\documentclass{article}

\usepackage[margin=0.75in]{geometry}
\usepackage{indentfirst} 
\usepackage[indent=0.25in]{parskip} 
\usepackage[none]{hyphenat} 
\usepackage{authblk} 
\usepackage{CJKutf8} 

\usepackage[cal=boondoxo,bb=libus,bbscaled=0.99,frak=euler]{mathalpha} 
\usepackage{amsmath,upgreek}

\usepackage{siunitx}
\DeclareSIUnit\fps{fps}

\usepackage{soul} 
\usepackage[dvipsnames]{xcolor} 
\usepackage{comment} 

\usepackage{csquotes}
\usepackage[backend=biber, sorting=none,style=nature, doi=false,url=true, eprint=false]{biblatex}  
\AtEveryBibitem{\clearfield{month}} 
\DeclareFieldFormat{pages}{\mkfirstpage{#1}}

\usepackage{hyperref}
\usepackage[capitalise, noabbrev]{cleveref} 
    \crefformat{equation}{#2Equation~#1#3}
    \crefformat{figure}{#2Figure~#1#3}
    \crefformat{table}{#2Table~#1#3}
    \crefrangeformat{figure}{Figures #3#1#4--#5#2#6}
\AtEveryBibitem{\clearfield{Month}}
    
\usepackage[labelfont=bf]{caption}
\usepackage{subcaption,graphicx,wrapfig,float}
\usepackage{wrapfig}

\usepackage{slashed}

\begin{document}
\title{How Easy Is It to Learn Motion Models from Widefield Fluorescence Single Particle Tracks?}
\author[1,2]{Zachary H. Hendrix}
\author[1,2]{Lance W.Q.~Xu~\begin{CJK*}{UTF8}{gbsn}(徐伟青)\end{CJK*}}
\author[1,2,3,$\dagger$]{Steve Press\'{e}}
\affil[1]{Center for Biological Physics,$ \  \ \: \, $Arizona State University, Tempe, AZ, USA}
\affil[2]{Department of Physics, $  \  \  \  \  \  \  \  \  \  \: \,  $Arizona State University, Tempe, AZ, USA}
\affil[3]{School of Molecular Sciences, $\, \  \: \, $Arizona State University, Tempe, AZ, USA}
\affil[$\dagger$]{Corresponding Author:  spresse@ASU.edu}
\date{\today}

\maketitle
\begin{abstract}

\noindent Motion models (\textit{i.e.}, transition probability densities) are often deduced from fluorescence widefield tracking experiments by analyzing single-particle trajectories post-processed from data. This analysis immediately raises the question: To what degree is our ability to learn motion models impacted by analyzing post-processed trajectories versus raw measurements? To answer this question, we mathematically formulate a data likelihood for diffraction-limited fluorescence widefield tracking experiments. In particular, we make the likelihood's dependence on the motion model versus the emission (or measurement) model explicit. The emission model describes how photons emitted by biomolecules are distributed in space according to the optical point spread function, with intensities subsequently integrated over a pixel, and convoluted with camera noise. Logic dictates that if the likelihood is primarily informed by the motion model, it should be straightforward to learn the motion model from the post-processed trajectory. Contrarily, if the majority of the likelihood is dominated by the emission model, the post-processed trajectory inferred from data is primarily informed by the emission model, and very little information on the motion model permeates into the post-processed trajectories analyzed downstream to learn motion models. Indeed, we find that for typical diffraction-limited fluorescence experiments, the emission model often robustly contributes $\approx\!99\%$ to the likelihood, leaving motion models to explain a meager $\approx\!1\%$ of the data. This result immediately casts doubt on our ability to reliably learn motion models from post-processed data, raising further questions on the significance of motion models learned thus far from post-processed single-particle trajectories from single-molecule widefield fluorescence tracking experiments.
\end{abstract}

\section*{Significance Statement}
\par We present a rigorous, physics-based statistical framework that clarifies the fundamental limitations of learning motion models---such as anomalous diffusion---from single particle trajectories acquired via widefield fluorescence microscopy. By deriving a likelihood that distinctly separates the contributions of the underlying motion and the emission (noise) model, we provide a clear understanding of the challenges inherent to these experimental conditions. Our framework offers a unifying explanation for conflicting reports in the literature and brings much-needed clarity to the interpretation of single-particle tracking data. This work has immediate repercussions on a very broad literature and even the future of anomalous motion as a model.

\begin{refsection}

\section*{Introduction}
\par The ability to deduce new physics from tracking experiments dates back to at least Robert Brown, who first inquired into the random motions of macroscopic pollen grains suspended in fluid\autocite{Brown1828}. This was followed by the work of Stokes\autocite{Stokes1851} and Fick\autocite{Fick1855}, who laid the foundation for Albert Einstein to formulate a physical theory invoking passive thermal fluctuations to explain such stochastic motions\autocite{Einstein1905}. Shortly thereafter, Norbert Wiener developed the first diffusive motion model. Concretely, he prescribed the mathematical form giving rise to the Gaussian transition probability density termed Brownian motion ($\mathrm{BM}$). In doing so, he developed a rigorous statistical framework encoding normal diffusion as mean-zero Gaussian displacements with stationary increments\autocite{Wiener1923DifferentialSpace}. As time passed, focus began shifting to anomalous diffusion\autocite{Kolmogorov1940, Montroll1965, Mandelbrot1968FBM}, describing diffusive motion with square displacement expectations, also termed mean squared displacements (MSDs), deviating from $\mathrm{BM}$'s linear time-dependence.

\par Initially, anomalous motion models were empirically inspired by macroscopic observations at reasonably high signal-to-noise ratio (SNR) regimes\autocite{Richardson1926, Mandelbrot1968FBM, Scher1975CTRW}. However, advances in instrumentation\autocite{SPTFLIM2023}, imaging techniques\autocite{sptPALM2008, MINFLUX2017, pMINFLUX2024}, alongside fluorescent labeling\autocite{Liu2023QDSVT} have created opportunities to look for anomalous diffusion at lower SNR down to the single-molecule regime\autocite{Saxton1997SPT, Metzler2009AnalysisSPT, 2009SPTI, 2009SPTII}. As a result, numerous methods have been designed with the intention of detecting anomalous diffusion from data\autocite{Manzo2021AnDiChallenge}, including methods to infer anomalous motion model parameters such as the anomalous exponent $\alpha$)\autocite{AnomalousUnicorns1992, AnomalousUnicorns2022, Krog2018fBmInference, LwInference, Metzler2022sBmVERSUSfBm, Gratin, DeepSPT2016DeepResidualLearning, DeepSPT2016ScalableBoosting, RANDI, TeamF, WadNET, TeamH, ELM, TeamJ2018Networks, TeamJ2019DeepLearning, TeamK, CONDOR, eRNN, TeamO2019, TeamO2020, TeamO2020Impact, Feng2024OOD, Zhang2025DynamicsEstimation}, methods to classify motion models\autocite{AnomalousUnicorns1992, AnomalousUnicorns2022, Krog2018fBmInference, LwInference, Metzler2022sBmVERSUSfBm, Gratin, DeepSPT2016DeepResidualLearning, DeepSPT2016ScalableBoosting, RANDI, TeamF, ELM, TeamJ2018Networks, TeamJ2019DeepLearning, TeamK, CONDOR, eRNN, TeamN2018, TeamN2019, TeamO2019, TeamO2020, TeamO2020Impact, Szarek2021Neural, Henrik2021DiffusionalFingerprinting, Feng2024OOD}, and tools for detecting changes in motion models along particle trajectories\autocite{Das2009HMM, RANDI, NoBIAS, TeamJ2018Networks, TeamJ2019DeepLearning, MunozGil2021Unsupervised, Metzler2023ML, Qu2024Semantic}.

\par Despite their different approaches, all published procedures used in learning motion models from data share a common trait: none directly consider raw experimental data (\textit{i.e.}, image stacks) as input. Instead, determining motion models relies on trajectories post-processed from data using single-particle tracking (SPT) software or statistical features calculated from these post-processed trajectories. Moreover, SPT algorithms themselves often assume a motion model as part of their inference. For example, through the cost function in the linear assignment problem\autocite{UTrack2008} or the dynamic model embedded in a Kalman filter\autocite{Chui2017, Presse2023}.

\par These considerations immediately raise the following question: is our ability to learn motion models impacted by analyzing trajectories post-processed from the data versus analyzing the raw measurements themselves? If nothing else, when ignored, static and dynamic localization errors, respectively, describing the instantaneous positional offset when localizing a particle and the motion-induced variance introduced when measuring its average position over a single, may generate non-linear lag-time dependence in MSD curves easily misinterpreted as anomalous diffusion for short trajectories\autocite{Berglund2010}. Although some studies have incorporated static localization errors into their analyses\autocite{Manzo2021AnDiChallenge}, SPT-derived trajectories are also susceptible to mislinking, which can further bias inferred motion models\autocite{Wolf2023, TARDIS}. Likewise, the misinterpretation of anomalous diffusion has already been identified from MSD analyses of diffusion in macrohomogeneous and microheterogeneous media over intermediate times\autocite{Berezhkovskii2014} and in ensemble-extracted diffusion time distributions\autocite{Kalwarczyk2007ApparentAnDi}.

\par To determine the effect of analyzing the post-processed trajectory in our ability to interpret the motion model of the resulting trajectory, we mathematically formulate a data likelihood for diffraction-limited fluorescence widefield tracking experiments.

\par
\begin{figure}[H]
    \centering
    \includegraphics[width=1.0\linewidth]{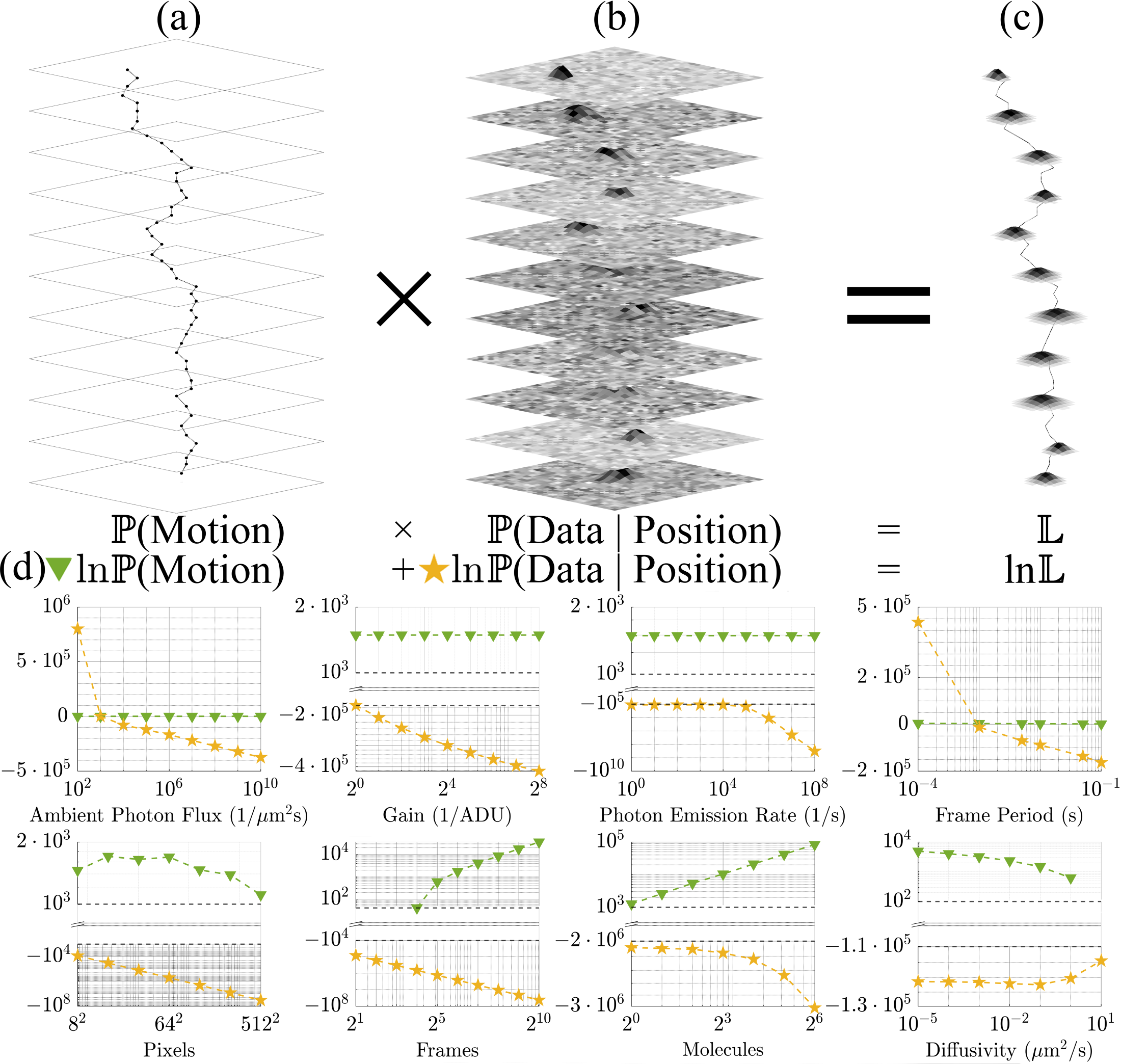}
    \caption{Our likelihood consists of emission and motion model portions. All data were collected for $\mathrm{BM}$ with the reference values in~\cref{Table(Parameters)} assigned where otherwise unspecified. 
(a) The motion model includes the transition probability, which explains how the current position is attained from prior locations. 
(b) The emission model describes how the detector output (ADUs) is related to the position (or multiple positions) of the particle attained within a frame.
(c) Likelihoods yield positions with some breadth of the likelihood function reflecting measurement error or finiteness of data.
(d) Numerical values for the emission and motion model portion of $\ln{\mathbb{L}}$ obtained for pure diffusion (\textit{i.e.}, $\mathrm{BM}$) as we vary one parameter at a time while the rest remain fixed at a standard value. Key to our argument here is the almost universally larger emission model magnitude. For a small number of frames $N\leq2^{3}$ or large diffusivities $\mathcal{D}\geq10\,\mathrm{\upmu m^{2}/s}$, the motion model obtains negative values, which cannot be shown alongside positive ones in log scale.}
    \label{Figure:Cartoon}
\end{figure}

 Following the logic of Hidden Markov Models (HMMs)\autocite{Presse2023}, it is convenient to express the likelihood as the product of two terms:
(i) the emission model prescribing how measurements are informed given the position of the molecule over each exposure, and 
(ii) the motion model describing the probability density according to which the current position is attained given previous positions. Written as such, the likelihood reads as follows:
\begin{equation}
    \mathbb{L}
\equiv
    \mathbb{P}\!\left(\text{Data, Position}\,\middle|\,\text{Motion}\right) 
= 
    \mathbb{P}\!\left(\text{Data}\,\middle|\,\text{Position}\right)
    \times
    \mathbb{P}\!\left(\text{Position}\,\middle|\,\text{Motion}\right),
\label{Equation:Likelihood(General)}
\end{equation}
with \(\mathbb{P}\!\left(\text{Data}\,\middle|\,\text{Position}\right)\) and \(\mathbb{P}\!\left(\text{Position}\,\middle|\,\text{Motion}\right)\) serving as our emission and motion models, respectively. The emission model, in particular, describes how photons emitted by fluorescently labeled particles, typically labeled biomolecules, are distributed across space following the optical point spread function with intensities integrated over each pixel area while convoluted with camera noise. A cartoon illustrating the breakdown of the likelihood is shown in \cref{Figure:Cartoon}.

In some literature, the likelihood is sometimes used to refer exclusively to the emission model\autocite{Lelek2021single, Wu2022Maximum}, thereby excluding the motion model from the definition of the likelihood. Here, our focus is really on comparing the relative importance of the motion and emission models. Naming conventions, as such, are unimportant so long as we are mathematically clear as to our definition of motion versus emission model.
\vspace*{-\baselineskip}

\par Logic dictates that if the contribution of the motion model is greater than that of the emission model, then the particle trajectories could easily be biased by the motion model assumed in the SPT tools. Conversely, if the contribution of the emission model exceeds that of the motion model, then the post-processed trajectory inferred from data is primarily informed by the emission model, and hence, with limited bias. However, this also indicates little information on the motion model permeates into the post-processed trajectories analyzed downstream to learn motion models. Leading with the conclusion and the results already hinted at within \cref{Figure:Cartoon}, we find that for typical diffraction-limited fluorescence experiments, the emission model is robustly two orders of magnitude greater in log space than the motion model.

\par The consequences of this result are far-reaching, underscoring the importance of working directly with raw data rather than post-processed trajectories. Our findings also call into question previously reported motion models derived from post-processed single-molecule widefield fluorescence trajectories. We argue that motion model classification depends critically on the quality of particle trajectories extracted from imaging data, making unbiased trajectory extraction essential for reliable classification. Moreover, the inability of many existing tools that we will explore to distinguish pure diffusion from anomalous diffusion highlights how little of the information on the motion model makes it through into the post-processed trajectory.

\section*{Results}
\par Here, we briefly highlight the logic of the presentation of our results.

\par In particular, we begin by demonstrating that we can reliably extract particle trajectories---with positions denoted by $\boldsymbol{R}_{1:N}$ expressing positions in $N$ frames---generated according to multiple motion models with negligible bias despite assuming a $\mathrm{BM}$ model in our likelihood. To do so, we quantify tracking error for data generated according to various motion models by calculating the percentage of ground truth positions circumscribed within the $98\%$ credible interval (CI) of inferred localizations assuming a $\mathrm{BM}$ model in our likelihood. Indeed, as shown in \cref{Figure:Results_Combined}, we accurately localized $99.9\%$ of true positions across data generated according to normal and anomalous motion models by analyzing the data with a $\mathrm{BM}$ model in the likelihood. The majority of the (typically large) error around each localization derived from SPT is induced by the breadth of the inherent emission model ascribed to the breadth of the point spread function itself, pixelization noise, finiteness of data, as well as detector noise.

\par After verifying trajectory extraction with $99.9\%$ accuracy across data generated using various motion models, we then quantify to what degree a trajectory learned is explained by the emission model versus its motion model contribution~\cref{Equation:Likelihood(General)}. To do so, we compare their relative probabilities and the shapes of their associated distributions. We simultaneously investigate the robustness of tracking in realistic SNR regimes across data generated according to various motion models. As later shown in~\cref{Figure:Results(Probability)}, the motion model contribution to the log likelihood never exceeds $10\%$, and, as we will see, often lies far below this, nearing $0.1\%$. This will help us quantitatively ascertain our conclusion that we robustly extract trajectories even for particles evolving according to anomalous motion models.

\par Given that the emission model often contributes $\approx\!\!\,99\%$ or more of the likelihood's probability logarithm in widefield fluorescence SPT, we will then explore the effect of static and dynamic/blurring localization errors on motion model classification. We will put our extracted particle trajectories, generated by both $\mathrm{BM}$ as well as anomalous diffusion, alongside corresponding ground truth trajectories and those extracted from TrackMate\autocite{Trackmate, Trackmate7} into software devised to classify motion models. We will see that tools used to classify motion models and infer parameters, \textit{CONDOR}\autocite{CONDOR} and \textit{AnomDiffDB}\autocite{TeamJ2018Networks, TeamJ2019DeepLearning}, only properly deduced pure diffusion (\textit{i.e.}, $\mathrm{BM}$) in $5/18 \ (28\%)$ of the trials employing $\mathrm{BM}$ trajectories. Indeed, we will demonstrate that static and dynamic localization errors create difficulties in accurately predicting the motion model, which is consistent with the notion that motion models contribute a small to insignificant portion of the likelihood.

\subsection*{Tracking Anomalous Diffusion with Negligible Bias}
\par
\begin{figure}[H]
    \centering
    \includegraphics[width=1\linewidth]{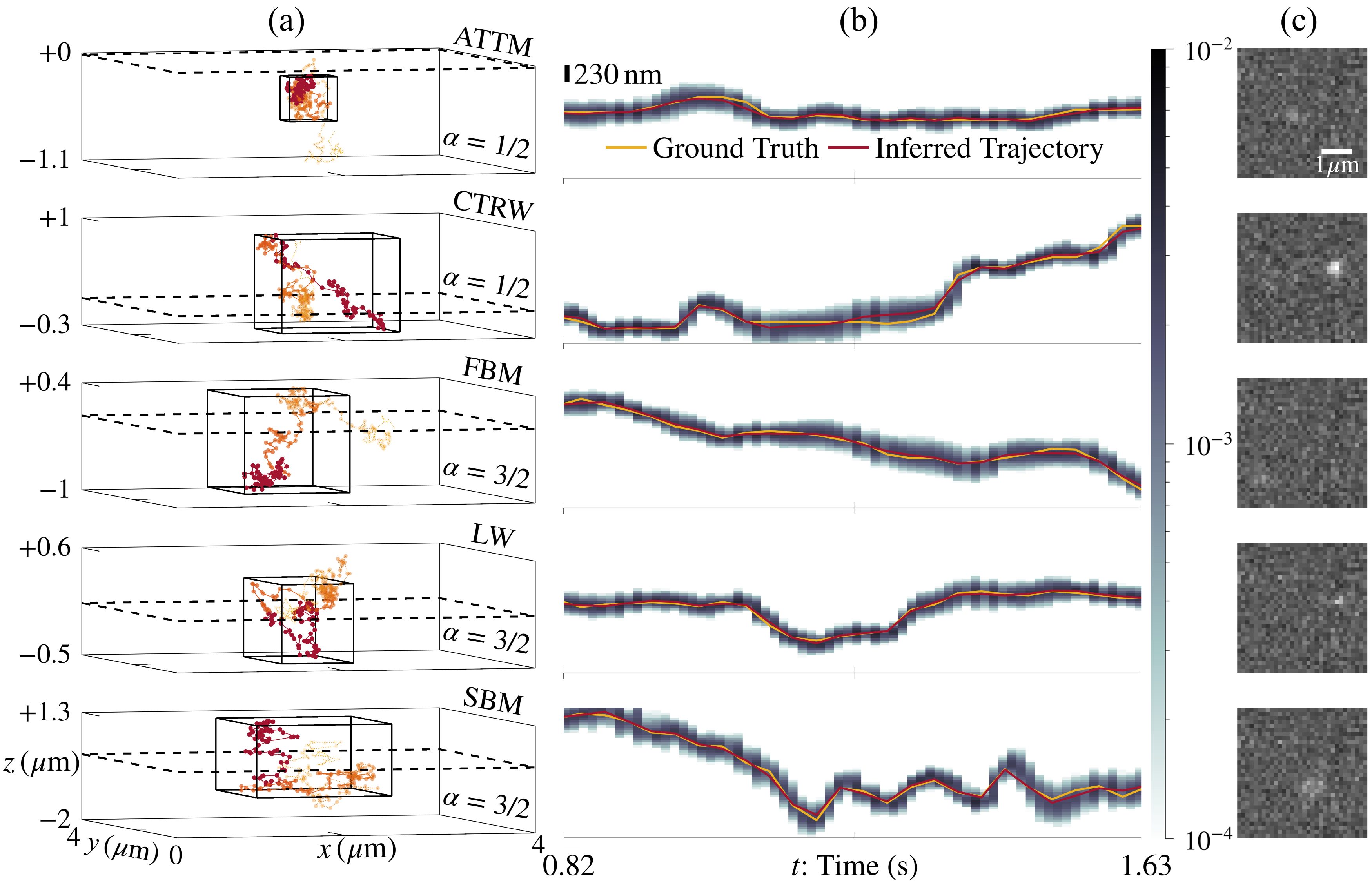}
\caption{A likelihood informed by the $\mathrm{BM}$ model accurately tracks particle positions with trajectories generated according to alternate motion models. 
(a) Inferred trajectories $\boldsymbol{R}_{1:N}$ are shown for each motion model with opacity, marker size, and color wavelength increasing with time. A dashed line marks the image plane, whereas the solid three-dimensional box encloses the samples shown in the central panel. 
(b) Three-dimensional trajectories along the $\hat{\boldsymbol{x}}$ direction are shown for ground truth (gold) and the mean MCMC sample (red); these trajectories are accompanied by a \qty{230}{\nano\m} scalebar and an associated shading that represents the $98\%$ CI obtained without burn-in over MCMC iterations $i\in[2,6]\cdot10^{3}$. 
(c) The final image $\boldsymbol{w}_{N}^{1:P}$ of each motion model is shown with a \qty{1}{\micro\m} scalebar; these frames have been transposed to align visually with central $x(t)$ plots. 
The generation of data for this figure is detailed in the \textbf{Forward Models for Data Acquisition} within the \textbf{Methods} section. A complete list of assigned measurement parameters is provided in \cref{Table(Parameters)}.
From top to bottom, the data associated with each panel is provided in {\bf Supplementary Image Stack 1:5 (.tiff)}, respectively.}
    \label{Figure:Results_Combined}
\end{figure}

 First, we set out to determine whether particles evolving according to anomalous diffusion motion models can be successfully tracked irrespective of which motion model is used in the likelihood to perform tracking. For this task, we evaluated the performance of \textit{BNP-Track}\autocite{Xu2024} and \textit{TrackMate}\autocite{Trackmate, Trackmate7} in the recovery of trajectories generated from particle tracks generated according to anomalous diffusion motion models. Below, \cref{Figure:Results_Combined} shows that we accurately tracked annealed time transient motion ($\mathrm{ATTM}$), a continuous time random walk ($\mathrm{CTRW}$), fractional Brownian motion ($\mathrm{FBM}$), a Levy walk ($\mathrm{LW}$), and scaled Brownian motion ($\mathrm{SBM}$) with high posterior probability. Quantitatively, from $13$ trajectories simulated according to motion models of anomalous diffusion, we recovered $649/650 \ (99.9\%)$ of true in-frame image-plane positions within the $98\%$ CI taken from samples $i\in[3,6]\cdot10^{3}$ to remove Markov Chain Monte Carlo (MCMC) burn-in. Our standard approach toward the removal of burn-in\autocite{Presse2023} is detailed in \textbf{Supplementary Information Part I: Burn-In Removal}, and the tracking of all $13$ anomalous and $2$ normal diffusive trajectories is presented in the \textbf{Supplementary Figures}.

\subsection*{Likelihood Contributors}
\par
\begin{figure}[H]
    \centering
\includegraphics[width=1\linewidth]{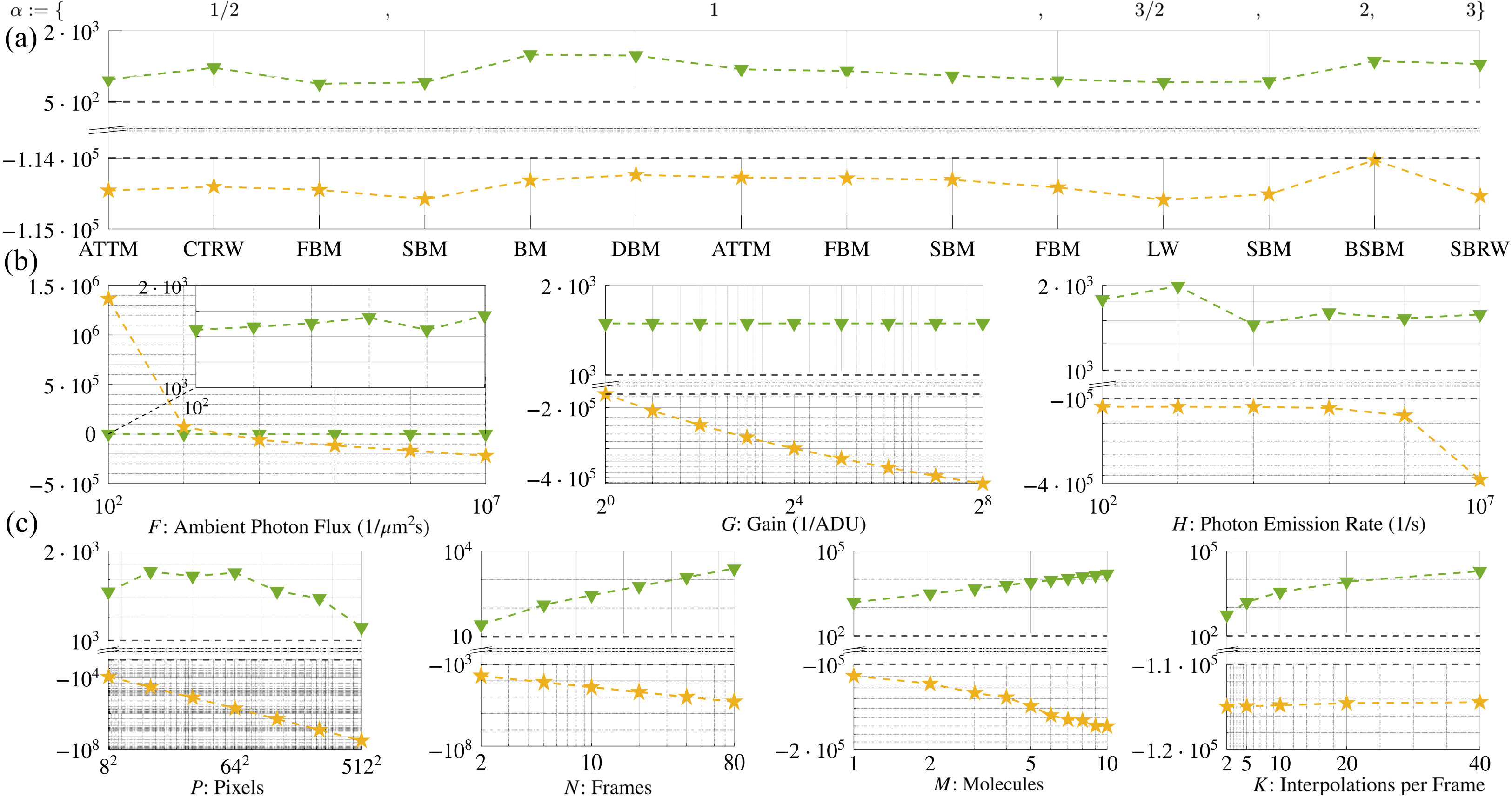} 
\caption{Our likelihood is not appreciably biased by motion models. Here, as in \cref{Figure:Cartoon}, we denote the probability logarithm of the emission model with a gold star and that of the motion model by a blue gradient. All data were collected for $\mathrm{DBM}$ with the reference values in~\cref{Table(Parameters)} assigned where otherwise unspecified.
(a) Mean probability logarithms for all motion models simulated. Mean values were computed along an interval of MCMC iterations selected to eliminate Monte Carlo burn-in: $i\in[2,6]\cdot10^{3}$. 
(b) Dependence of probability logarithms on emission parameters. 
(c) Dependence of probability logarithms on dimensionless parameters.}
\label{Figure:Results(Probability)}
\end{figure}

 Having demonstrated accurate tracking using a likelihood with a $\mathrm{BM}$ model irrespective of motion model and anomalous exponent of particles giving rise to the data, we now address the extent to which motion and emission models explain observations. We do so by assessing the numerical contribution of each to our likelihood under realistic imaging scenarios. To avoid numerical underflow, our investigation into the robustness of tracking is confined to log space. Overall, as shown below in \cref{Figure:Results(Probability)}, we find that our emission model explains a majority of the data in all realistic imaging scenarios over the contribution of the motion model to the likelihood, irrespective of which motion model was used to generate the data. That is, how the photons are distributed across pixels and how the signal is convoluted with detector noise is far more important in tracking than any \textit{a priori} assumption used in the likelihood on the motion model. From this, we conclude that motion models do not appreciably bias trajectories inferred using a $\mathrm{BM}$ model ~\cref{Equation:Likelihood(General)}.
\vspace*{-\baselineskip}

\par As captured by~\cref{Figure:Results(Probability)} (a), the substantial numerical difference separating the probability logarithms of the emission and motion model contributions to the likelihood span five orders of magnitude for data generated from all motion models considered here but was maximal for $\mathrm{BM}$ and $\mathrm{DBM}$ with $\mathcal{D}:=$\qty{0.05}{\micro\m^2/\s}, ballistic $\mathrm{SBM}$, and a superballistic  random walk ($\mathrm{SBRW}$). What we find is that our emission model predominantly determines the inference of a trajectory from observations, regardless of which motion model generated the data. That is, motion models do not appreciably bias trajectory inference. This result confirms that particle trajectories can be robustly extracted across motion models, whilst verifying that the optics of an imaging system with detector noise, as informed by an emission model, provide much more substantive information than displacements statistics (\textit{i.e.}, models of transition probability), as informed by a motion model.

\par Given that our log likelihood is negligibly informed by the motion model and thus can reliably learn trajectories of particles generated with non-$\mathrm{BM}$ models, we now investigate the probabilistic weights of our log likelihood. As expected and shown in \cref{Figure:Results(Probability)} (b:c), the emission model's probability logarithm remains at least one order of magnitude larger than that of the motion model, regardless of which parameters are individually varied. Thus, our emission model remains the principal driver of trajectory inference, given that the image plane has sufficient background illumination.

\par In principle, there exists a single exception to the data-driven inference condition we have demonstrated in the preceding paragraph: an arbitrarily large number of positions could be interpolated between frames such that the motion model's contribution could eventually dominate the likelihood---a consequence of shortening the lag time $\mathit{\Delta}t$ between successive time intervals. In practice, however, interpolating multiple intraframe positions is uncommon, as these positions become highly uncertain and significantly slow down inference; indeed, computation time scales exponentially with $K$.

\subsection*{Motion Model Classification}
\par Having demonstrated robust tracking irrespective of the motion model generating diffusive trajectories in realistic SNR regimes, we now investigate tools for motion model classification by considering whether existing methods\autocite{CONDOR, TeamJ2018Networks, TeamJ2019DeepLearning} can detect motion models for data generated according to anomalous diffusion ($\alpha\ne1$) and $\mathrm{BM}$ provided we input to these methods: i) ground truth trajectories; and ii) those inferred by using both TrackMate\autocite{Trackmate, Trackmate7}; and iii) \textit{BNP-Track}\autocite{Xu2024} to analyze the data. In doing so, we also surveyed the usability of other methods serving as additional submissions to the Anomalous Diffusion (\textit{AnDi}) Challenge\autocite{Manzo2021AnDiChallenge}. The majority of software developed in Python\autocite{AnomalousUnicorns1992, AnomalousUnicorns2022, Gratin, DeepSPT2016DeepResidualLearning, DeepSPT2016ScalableBoosting, RANDI, TeamF, WadNET, TeamJ2018Networks, TeamJ2019DeepLearning, TeamK, eRNN, TeamN2018, TeamN2019, TeamJ2019DeepLearning, TeamO2019, TeamO2020, TeamO2020Impact} have become unusable due to library deprecations and package incompatibilities five years following the Challenge. Others, like \textit{NOBIAS}\autocite{NoBIAS}, failed to execute with the data provided in its own repository, as seen in the Table 1 of \textbf{Supplementary Tables}.

\par Thus, for motion model classification and parameter inference from ground truth and post-processed trajectories, we selected functioning top contenders from the Anomalous Diffusion Challenge\autocite{Manzo2021AnDiChallenge}---\textit{CONDOR}\autocite{CONDOR} and \textit{AnomDiffDB}\autocite{TeamJ2019DeepLearning}. Despite its high accuracy in the classification challenge\autocite{Manzo2021AnDiChallenge}, \textit{AnomDiffDB}\autocite{TeamJ2018Networks, TeamJ2019DeepLearning} proved less flexible in practice; its convolution neural network architecture, requiring no less than $N=100$ frames for analysis, may be incompatible with realistic acquisition constraints of experimental data, but it provided a reliable benchmark. While \textit{AnomDiffDB} can only classify three motion models $($\textit{i.e.}, $\mathit{\Theta}=\{\mathrm{BM}, \, \mathrm{FBM}, \, \mathrm{CTRW}\}$$)$, \textit{CONDOR}\autocite{CONDOR}--which draws inferences from calculated trajectory features, \textit{e}.\textit{g}., mean squared displacement (MSD) and power spectral density (PSD) analyses--can classify all anomalous motion models featured in the anomalous diffusion challenge\autocite{Manzo2021AnDiChallenge}, \textit{i.e.}, \\ $\mathit{\Theta}=\{\text{ATTM},\,\text{CTRW},\,\text{FBM},\,\text{LW},\,\text{SBM}\}$.

\par First, in order to validate classification tools (\textit{AnomDiffDB}\autocite{TeamJ2018Networks, TeamJ2019DeepLearning} and \textit{CONDOR}\autocite{CONDOR}) from perfect, noiseless data, we forewent static and dynamic localization errors altogether by inputting ground truth $\mathrm{BM}$ trajectories into motion model classification software. Then, to demonstrate the necessity of incorporating an emission model to account for static and dynamic localization errors in trajectories recovered from analyzing data, we used both \textit{BNP-Track}\autocite{Xu2024} and \textit{TrackMate}\autocite{Trackmate, Trackmate7} to infer these $\mathrm{BM}$ trajectories from the data before performing motion model classification.
\begin{table}[H]
    \centering
    \begin{tabular}{|c|c|c|c|} \hline 
           $\mathcal{D} \  [\upmu\mathrm{m^2/s}]$&$\text{Trajectory}$&\textit{AnomDiffDB}&  \textit{CONDOR} \\ \hline
 & $\text{Ground Truth}$& $\text{FBM} \ (73\%)$&$ \ \ \, \text{FBM}(\hat{\alpha} = 1.08)$\\ $1$& $\text{BNP-Track}$& $\text{FBM} \ (78\%)$&$\text{ATTM}(\hat{\alpha} = 0.95)$\\
 & $\text{TrackMate}$& $\text{FBM} \ (75\%)$&$ \ \ \, \text{SBM}(\hat{\alpha} = 1.13)$\\ \hline
 & $\text{Ground Truth}$& $\ \,\text{BM} \ (72\%)$&$ \ \ \, \text{FBM}(\hat{\alpha} = 1.08)$\\
 $5$& $\text{BNP-Track}$& $\text{FBM} \ (74\%)$&$\text{ATTM}(\hat{\alpha} = 1.00)$\\
 & $\text{TrackMate}$& $\ \, \text{BM} \ (60\%)$&$\text{ATTM}(\hat{\alpha} = 0.55)$\\ \hline
 & $\text{Ground Truth}$& $\ \, \text{BM} \ (47\%)$&$ \ \ \, \text{FBM}(\hat{\alpha} = 1.08)$\\
 $10$& $\text{BNP-Track}$& $\text{FBM} \ (64\%)$&$\text{ATTM}(\hat{\alpha} = 0.95)$\\
 & $\text{TrackMate}$& $\ \,\text{BM} \ (72\%)$&$\text{ATTM}(\hat{\alpha} = 0.95)$\\ \hline
    \end{tabular}
\caption{Trajectory inference biases motion model classifications. 
Here, we generated $\mathrm{BM}$ ground truth trajectories and their synthetically imaged data with the diffusivities shown in the left column. All data were generated with the reference values in \cref{Table(Parameters)} assigned to parameters.  
Using \textit{BNP-Track}\autocite{Xu2024} and \textit{TrackMate}\autocite{Trackmate, Trackmate7}, we inferred these $\mathrm{BM}$ trajectories from their image stacks. 
We then independently input ground truth and each set of inferred trajectories into state-of-the-art software \textit{AnomDiffDB}\autocite{TeamJ2018Networks, TeamJ2019DeepLearning} and \textit{CONDOR}\autocite{CONDOR} to perform motion model classification. In the \textit{AnomDiffDB} column, we show the motion model inferred by this tool and the confidence with which it was predicted. In the \textit{CONDOR} column, we display which motion model this tool classified and the anomalous exponent it estimated $\hat{\alpha}$.
 All \textit{CONDOR } classifications from inferred $\mathrm{BM}$ trajectories contradict those drawn from ground truth with at least $5\% $ error between the estimated $\hat{\alpha}$ and the true ``anomalous" exponent $(\alpha=1)$ generating the data, whereas $78\%$ of \textit{AnomDiffDB}'s classifications matched across trajectory extraction methods.}
    \label{Table(AnomalousInference)}
\end{table}
\vspace*{-\baselineskip}

\par On the one hand, \cref{Table(AnomalousInference)} shows that \textit{CONDOR} identically classified motion models from ground truth $\mathrm{BM}$ trajectories within $7.5\%$ error on estimates $\hat{\alpha}$ with respect to the true ``anomalous" exponent $(\alpha=1)$ generating the data, highlighting the consistent precision offered by feature-based methods given perfect, noiseless trajectories generated using identical pseudorandom number generator seeds. Nevertheless, this feature-based method clearly biased classifications from inferred $\mathrm{BM}$ trajectories, classifying $5/6 \ (83\%)$ of such trajectories as $\hat{\mathit{\Theta}}=\mathrm{ATTM}$, which consists of localized patches of $\mathrm{BM}$ each with their own diffusivity. \textit{CONDOR}'s obvious bias towards $\mathrm{ATTM}$ implies that static and dynamic localization errors manifest as apparent changes in $\mathcal{D}$ over the duration of pure diffusion, as this method did not predict $\mathrm{ATTM}$ for any ground truth $\mathrm{BM}$ trajectory. \textit{BNP-Track} reconstructed $\mathrm{BM}$ trajectories in a manner that led \textit{CONDOR} to accurately classify the $\mathcal{D}:= 5\,\mathrm{\upmu m^{2}/s}$ trajectory as pure diffusion [\textit{i.e.}, $\text{ATTM}(\hat{\alpha}=1)$] but skewed the remaining $\mathcal{D}:=\{ 1,10\}\,\mathrm{\upmu m^{2}/s}$  trajectories toward anomalous subdiffusion such that \textit{CONDOR} predicted $\hat{\alpha}=0.95$ for both. For $\mathrm{BM}$ trajectories inferred from \textit{TrackMate}, however, \textit{CONDOR} inconsistently classified motion models and inferred anomalous exponents. These results highlight the challenges in attempting trajectory classification starting from post-processed data.

\par On the other hand, \cref{Table(AnomalousInference)} shows that \textit{AnomDiffDB} predicted a different combination of motion model and anomalous exponent for each ground truth $\mathrm{BM}$ trajectory, becoming less certain but more accurate with greater diffusivities: the slow $\mathrm{BM}$ trajectory was misclassified as anomalous $\hat{\mathit{\Theta}}=\mathrm{FBM}$ with $73\%$ confidence, whereas the fastest $\mathrm{BM}$ trajectory was correctly identified as pure diffusion but with only $47\%$ certainty. For trajectories extracted by \textit{BNP-Track}, however, \textit{AnomDiffDB} never made an accurate prediction but preserved the anticorrelation observed between certainty and diffusivity, losing $14\%$ confidence between the slowest and fastest $\mathrm{BM}$ trajectories. As for \textit{TrackMate} trajectories, \textit{AnomDiffDB}'s accuracy was again correlated with diffusivity, accurately classifying $2/3 \ (67\%)$ $\mathrm{BM}$ trajectories as pure diffusion; interestingly, only $3\%$ confidence was lost between the slow and fast $\mathrm{BM}$ trajectory. Nonetheless, all of these outcomes confirm that motion model classifications are biased by trajectory inference; furthermore, the $25\%$ confidence increase observed for predictions made from the fast $\mathrm{BM}$ trajectory extracted by \textit{TrackMate}---versus ground truth---hint at the fact that modular tracking algorithms, which decompose tracking into sequential, independently optimized modules (of particle determination, localization, and linking) rather than performing unified global optimization, may reconstruct trajectories with bias towards $\mathrm{BM}$. 

\par The findings above underscore the difficulty of motion model classification in the presence of static and dynamic localization errors, highlighting the importance of incorporating the statistical information made available by considering a carefully calibrated emission model. Even before accounting for static and dynamic localization noise, existing classifiers\autocite{CONDOR, TeamJ2018Networks, TeamJ2019DeepLearning} are heavily prone to misinterpreting pure diffusion as anomalous diffusion.

\section*{Methods}
\subsection*{The Likelihood of Widefield Fluorescence SPT}
\par As before, we write down our general likelihood as the product of an emission model $\mathbb{P}\!\left(\text{Data} \, \middle| \, \text{Position} \right)$ with a motion model $\mathbb{P}\!\left(\text{Position}\,\middle|\,\text{Motion}\right)$:
\[
    \mathbb{L}
=  
    \mathbb{P}\!\left(\text{Data} \,\middle|\, \text{Position} \right)
\times
    \mathbb{P}\!\left(\text{Position}\,\middle|\,\text{Motion}\right).
\]
We explicitly formulate both components for widefield fluorescence SPT, capturing a particle diffusing in three dimensions over $N$ frames, interpolated at $K$ positions within a frame, and imaged over $P$ pixels.

\par Before deriving the motion model, we must first address the particle's initial localization. To accurately model the localization error attributed to the optical imaging system's point spread function $(\text{PSF})$, we consider the initial localization of a single particle in frame $n=1$ at position $\boldsymbol{R}_{n=1}$ to be approximated by a Gaussian distribution centered at the optical axis $\boldsymbol{\mu}=(\mu_{x},\mu_{y},0)$ with variance $\boldsymbol{\sigma}\odot\boldsymbol{\sigma}=(\sigma_{xy}^{2},\sigma_{xy}^{2},\sigma_{z}^{2})$, where $\odot$ denotes the Hadamard product, $\sigma_{xy}$ is the PSF's lateral width, and $\sigma_{z}$ is the PSF's axial width. Hence, the probability of observing the particle at this first position is
\begin{equation}
\mathbb{P}(\boldsymbol{R}_{1})
=
\left({\sigma_{xy}^{2}\sigma_{z}\sqrt{2\uppi}^{3}}\right)^{-1}
\exp\!{\left[-\frac{(X_{1}-\mu_{x})^{2}+(Y_{1}-\mu_{y})^{2}}{2\sigma_{xy}^{2}}-\frac{Z_{1}^{2}}{2\sigma_{z}^{2}}\right]
}.
\label{Equation:InitialLocalization}
\end{equation}

\par With the initial localization addressed, we now model the particle's evolving position as Brownian (\textit{i.e.}, statistically independent, stationary, Gaussian) transitions with diffusivity $\mathcal{D}$, interpolated at $K$ positions between each of the $N$ observed frames to give us the ability, if we so choose, to deduce dynamics on timescales exceeding data acquisition\autocite{Presse2021HMMs}. By interpolating positions between frames, we may resolve short-timescale dynamics and ultimately introduce a knob allowing us to test the reliability of positions inferred between frames (``intraframe" motion). Our motion model (\textit{i.e.}, transition probability density), codifying the probability of observing the particle at all successive positions $\boldsymbol{R}_{2:N}^{1:K}$ under a $\mathrm{BM}$ model, therefore reads
\begin{equation}
\mathbb{P}\!\left(\text{Position}\,\middle|\,\text{Motion}\right)
=
\mathrm{exp}\!
\left[
    {-
        \sum\limits_{k = 1}^{K}
        \sum\limits_{n = 2}^{N}
            \frac{\left(\mathit{\Delta}R_{n}^{k}\right)^{2}}
            {4\mathcal{D}\mathit{\Delta}t_{n}^{k}}}
            \right]
    \prod_{k=1}^{K}
        \prod_{n=2}^{N}
        \left(4\uppi\mathcal{D}\mathit{\Delta}{t}_{n}^{k}\right)^{-3/2},
        \label{Equation:MotionModel}
\end{equation}
where $\mathit{\Delta}{t}_{n}^{k}$ is the duration between the $k$\textsuperscript{th} and preceding interpolation within frame $n$, and $\mathit{\Delta}R_{n}^{k}$ is the corresponding Euclidean distance. In \textbf{Supplementary Information Part II: Alternate Motion Models}, we specify $\mathbb{P}\left(\text{Position}\,\middle|\,\text{Motion}\right)$ for motion models $\mathit{\Theta}=\{\mathrm{DBM},\mathrm{FBM}$\autocite{Krog2018fBmInference}$,\mathrm{LW}$\autocite{LwInference}$,\mathrm{SBM}$\autocite{Metzler2022sBmVERSUSfBm}$\}$.

\par While our motion model, \cref{Equation:MotionModel}, governs the evolution of a particle's position over time, we now describe the emission distribution capturing both static and dynamic localization noise. As photon detection occurs across all $P$ pixels within each of the $N$ frames, its statistics are shaped by the particle's photon emission rate $H$, the ambient photon flux $F$ contributing background illumination, and the optical imaging system's $\text{PSF}$. Mathematically, the expected signal $u_{n}^{p}$, which determines the mean photon count incident upon the $p$\textsuperscript{th} pixel in the $n$\textsuperscript{th} frame in analog-to-digital units (ADU), captures both static and dynamic localization errors by integrating photon emissions over the pixel area and exposure period, convolved with the point spread function centered at the particle's position:
\begin{equation}
u_{n}^{p}
= 
F A^{p}\tau
+
\, \hspace{-1em} 
\int\limits_{t_{n-1} \ }^{ \ t_{n}}
\ \hspace{-1em} 
    \mathrm{d}{t} \!\!\!\:
    \iint\limits_{A^{p} \, }
        {\mathrm{d}{A^{p}}
        \left[
            H \: 
            \mathrm{PSF}(x^{p},y^{p} \, ; \boldsymbol{R}_{n})
        \right]
        }.
\label{Equation:Expectation}
\end{equation}
In \cref{Equation:Expectation}, background illumination arises from spatiotemporally compounding $F$ over the $p$\textsuperscript{th} pixel's area $A^{p}
\equiv
\int_{\underline{y}^{p}}^{\bar{y}^{p}}
    \mathrm{d}y
    \left(
        \int_{\underline{x}^{p}}^{\bar{x}^{p}}
        \mathrm{d}x
    \right)$ and over the $n$\textsuperscript{th} frame's effective exposure period $\tau\equiv\int_{t_{n-1}}^{t_n}\mathrm{d}t - t_{\text{dead}}$, given a detector's dead time $t_{\text{dead}}$\autocite{Presse2020SinglePhoton, Presse2024FluorescenceMicroscopy, Xu2024}. Here, $\underline{x}$ represents the $x$-interval's lower bound, and $\bar{x}$ its upper bound.

\par Static and dynamic localization noise introduce probabilistic and deterministic errors in measured positions induced by low photon counts and the motion of objects within the field of view relative to the detector, respectively, whereas the detector recording each frame additionally introduces a stochastic measurement process that depends on its signal amplification process. Here, we formulate the measurement process of an EMCCD operated with high electron-multiplying (EM) gain; in the \textbf{Supplementary Information Part III: Alternate Emission Models}, however, we derive the Poisson-Gamma-Normal (PGN) model\autocite{Hirsch2013EMCCD, Ryan2021EMCCD} for arbitrary EM gains as well as the analogous measurement process given a detector with sCMOS architecture\autocite{Huang2011Localization, Huang2013, Liu2017sCMOS, Mandracchia2020sCMOS}. Independently arriving photons absorbed within the detector's exposure period generate photoelectrons in each pixel with Poisson statistics\autocite{Hirsch2013EMCCD} arising from the quantum nature of light and facilitated by the photoelectric effect. As such, each photon can excite no more than a single conduction-band electron with probability dictated by the quantum efficiency $\beta$, tending to take values of $\beta\!>\!90\%$ in EMCCDs\autocite{GammaEMCCD}. Photoelectrons undergo impact-ionization in the EMCCD's multiplication register, exciting secondary electrons in the process. Since impact-ionization introduces small gains in electron $(\mathrm{e^{-}})$ counts across the individual high-voltage wells comprising the multiplication register, the process is mathematically equivalent to a long cascade of Bernoulli or Poisson branchings whose variance grows with electron counts; thus, compounding this gain over hundreds of wells yields a Gamma distribution characterizing the electron count exiting the register, transforming the expected analog intensity \cref{Equation:Expectation} into a measured count $w_{n}^{p}$ (in ADU) well modeled by
\begin{equation}
    w_{n}^{p}\sim\text{Gamma}\!\left(\beta u_{n}^{p}/2,\, 2G/\varphi\right),
    \label{Equation:Measurement}
\end{equation}
where the shape parameter $(\beta u_{n}^{p}/2)$ represents half the expected incident photons exciting conduction electrons after accounting for quantum losses. The scale parameter $(2G/\varphi)$ is just twice the user set electron-multiplication gain $G$\autocite{Hirsch2013EMCCD} converted into ADU by the conversion factor (\textit{i.e.}, calibration parameter) $\varphi$ carrying units of $\mathrm{e^{-}/ADU}$\autocite{Ryan2021EMCCD}. This model, \cref{Equation:Measurement}, is particularly accurate for relatively large gains $G\geq210$\autocite{Hirsch2013EMCCD} that drive the gamma multiplication statistics to dominate over the Gaussian readout noise completely. Compounding~\cref{Equation:Measurement} over all $P$ pixels and $N$ frames and accounting for the particle's initial position, the emission portion of the likelihood then reads
\begin{equation}
\begin{split}
    \mathbb{P}(\text{Data}\,|\,\text{Position})
& = 
    \left({\sigma_{xy}^{2}\sigma_{z}\sqrt{2\uppi}^{3}}\right)^{-1}
    \exp\!{\left[-\frac{(X_{1}-\mu_{x})^{2}+(Y_{1}-\mu_{y})^{2}}{2\sigma_{xy}^{2}}-\frac{Z_{1}^{2}}{2\sigma_{z}^{2}}\right]}
\\ & \times 
    {\mathrm{exp}\!
    \left({-\sum\limits_{p=1}^{P}\sum\limits_{n=1}^{N}
        \frac{w_{n}^{p}}{2G/\varphi}}
    \right)}
    \prod_{p = 1}^{P}
    \prod_{n = 1}^{N}
        {\left(w_{n}^{p}\right)
        ^{\frac{\beta u_{n}^{p}}{2} - 1}}
        \left[
             \left(\dfrac{2G}{\varphi}\right)^{\!\!\frac{\beta u_{n}^{p}}{2}}
            \!\!\!\!\,
            \Gamma\!\left(\dfrac{\beta u_{n} ^{p}}{2}\right)
        \right]^{-1}.
\label{Equation:EmissionModel}
\end{split}
\end{equation}

\par The full likelihood, obtained by substituting the emission model \cref{Equation:EmissionModel} and motion model \cref{Equation:MotionModel} into \cref{Equation:Likelihood(General)}, now reads
\begin{subequations}
\begin{align}
\mathbb{L} 
&= 
\left({\sigma_{xy}^{2}\sigma_{z}\sqrt{2\uppi}^{3}}\right)^{-1}
\exp\!{\left[-\frac{(X_{1}-\mu_{x})^{2}+(Y_{1}-\mu_{y})^{2}}{2\sigma_{xy}^{2}}-\frac{Z_{1}^{2}}{2\sigma_{z}^{2}}\right]
}
\\ & \times
{\mathrm{exp}\!
\left(
    {-\sum\limits_{p=1}^{P}\sum\limits_{n=1}^{N}
        \frac{w_{n}^{p}}{2G/\varphi}}
\right)
        }
\prod_{p = 1}^{P}
\prod_{n = 1}^{N}
        {\left(
                w_{n}^{p}
            \right)^
            {\frac{\beta u_{n}^{p}}{2} - 1}}
    \left[
        \left(\dfrac{2G}{\varphi}\right)^{\!\!\frac{\beta u_{n}^{p}}{2}}
        \!\!\!\! \,
        \Gamma\!\left(\dfrac{\beta u_{n} ^{p}}{2}\right)
    \right]
    ^ {-1}
\\ & \times
\mathrm{exp}\!
\left[
    {-
        \sum\limits_{k = 1}^{K}
        \sum\limits_{n = 2}^{N}
            \frac{\left(\mathit{\Delta}R_{n}^{k}\right)^{2}}
            {4\mathcal{D}\mathit{\Delta}t_{n}^{k}}}
            \right]
    \prod_{k=1}^{K}
        \prod_{n=2}^{N}
        \left(4\uppi\mathcal{D}\mathit{\Delta}{t}_{n}^{k}\right)^{-3/2}.
\end{align}
\label{Equation:Likelihood}
\end{subequations}
Our likelihood, \cref{Equation:Likelihood}, characterizes the evolution of a particle's position over time through a motion model (c) independent from the emission model characterizing both optical limitations (a) and photon detection with measurement degradation (b). Through this likelihood, we examine how much a trajectory can be explained through the emission and motion model contributions after accurately tracking particles, irrespective of which motion model generated the data. To circumvent numerical underflow, we constrain our investigation of statistical information to log space and present the logarithm of \cref{Equation:Likelihood} as the sum of its components: $\ln{\mathbb{L}}=\ln\mathbb{P}\left(\text{Data}\, | \, \text{Position}\right)+\ln\mathbb{P}\left(\text{Position}\,\middle|\,\text{Motion}\right)$ as follows
\begin{subequations}
\begin{align}
\ln \mathbb{L} = &\ 
- 
\left[
\ln\!{\left[(2\uppi)^{3/2}\sigma_{xy}^{2}\sigma_{z}\right]}
    + \frac{(X_1 - \mu_x)^2 + (Y_1 - \mu_y)^2}{2\sigma_{xy}^2}
    + \frac{Z_1^2}{2\sigma_z^2} 
\right]
\\
&\ 
+ 
    \sum_{p=1}^{P} \sum_{n=1}^{N} 
    \left[
       -\frac{w_{n}^{p}}{2G/\varphi} 
        + \left(\frac{\beta u_{n}^{p}}{2}-1\right) \ln{w_{n}^{p}}
        - \frac{\beta u_{n}^{p}}{2} \ln{\dfrac{2G}{\varphi}} 
        - \ln\Gamma\frac{\beta u_{n}^{p}}{2}
    \right]
\\
&\ 
- \sum_{k=1}^{K} \sum_{n=2}^{N} 
    \left[
        \frac{(\mathit{\Delta} R_n^k)^2}{4\mathcal{D} \mathit{\Delta} t_{n}^{k}}
        + \frac{3}{2} \ln \left(4\uppi \mathcal{D} \mathit{\Delta} t_{n}^{k}\right) 
    \right].
\end{align}
\label{Equation:logLikelihood}
\end{subequations}
We then compare the probability logarithms of the emission model \cref{Equation:logLikelihood} (a:b) and the motion model \cref{Equation:logLikelihood} (c) to determine which contributes most. To evaluate the relative probabilistic weights of our likelihood in different SNR regimes, we also performed independent parameter variation.

\subsection*{Forward Models for Data Acquisition}
\par To help compare the magnitude of different contributors to our likelihood~\cref{Equation:Likelihood}, we synthesized data and assessed the relative contributions to our likelihood using known ground truths. For clarity, a brief forward model for generating data according to various motion models is presented here. Additionally, the data and code associated with this study have been made publicly available in the \textbf{Code \& Data Availability} section.
\par The trajectories shown in \cref{Figure:Results_Combined} were synthesized using the AnDi Challenge repository\autocite{Manzo2021AnDiChallenge, AnDiRepository}. For all \textit{single} particle (\textit{i.e.}, $M=1$) data, position three-vectors $\boldsymbol{R}$ were simulated over time $\{t_{n,k}\}$ discretized over $N$ frames and interpolated $K$ times between each frame. While the initial position $\boldsymbol{R}_{1}$ was drawn identically from the optical axis $\boldsymbol{\mu}$ and localization variance $\boldsymbol{\sigma}\odot\boldsymbol{\sigma}$ for all data, trajectories exhibiting persistent diffusion [\textit{i.e.}, $\mathrm{LW}$ and $\mathrm{SBM}$($\alpha>1$)] were normalized and uniformly rescaled before applying translations to center them within the image plane.

\subsubsection*{Generating Diffusive Trajectories}
\par Here, we succinctly introduce how particle trajectories in \cref{Figure:Results_Combined} were generated through various motion models after drawing an initial position $\boldsymbol{R}_{1}$ from \cref{Equation:InitialLocalization}. To avoid confusion with frame indices $n$ and indices for interpolated times between frames $k$, we introduce $\ell=1:L$ as the index for simulation times; thus, the models below populate the successive positions $\boldsymbol{R}_{2:L}$. \textit{Nota bene}: trajectories simulated using the \textit{AnDi Challenge} repository\autocite{AnDiRepository} are initialized at the spatial origin $(\boldsymbol{R}_{1}=\boldsymbol{0})$ by default, but such trajectories are translated near the optical center in the process of cropping the image frame to a region of interest (ROI) spanning $32\times32$ pixels.
\\ \vspace*{-\baselineskip}

\noindent \fbox{\parbox{\linewidth}{
\textbf{BM}: In three-dimensions, a Brownian walker's successive $\ell$\textsuperscript{th} position $\boldsymbol{R}_{\ell}$ is drawn from the multivariate normal (\textit{i.e.}, Gaussian) distribution $\mathbb{N}(\boldsymbol{\mu}, \boldsymbol{\Sigma}_{\ell})$ whose mean is the preceding position, $\boldsymbol{\mu}=\boldsymbol{R}_{\ell - 1}$, and whose isotropic covariance matrix is $\boldsymbol{\Sigma}_{\ell}=2\mathbb{I}\mathcal{D}\mathit{\Delta}{t}_{\ell}$:
\begin{equation}
    \left.\boldsymbol{R}_{\ell} \,\middle|\, 
    \boldsymbol{R}_{\ell-1}\right.
\sim
    {\mathbb{N}(\boldsymbol{R}_{\ell-1},2\mathbb{I}\mathcal{D}\mathit{\Delta}{t}_{\ell})},
\label{Equation:BM}
\end{equation}
where $\mathcal{D}$ is the particle's diffusivity, $\mathit{\Delta}{t}_{\ell} \equiv t_{\ell} - t_{\ell-1}$ is the lag-time between steps, and $\mathbb{I}$ is the $(3\times3)$ identity matrix.
}}

\noindent \fbox{\parbox{\linewidth}{
\textbf{DBM}: In diffusion with drift (\textit{i.e.}, $\text{DBM}$), a random walker's position accrues deterministic flow $(\boldsymbol{v}\mathit{\Delta}{t}_{\ell})$ alongside the stochastic Wiener process of $\mathrm{BM}$ in \cref{Equation:BM}:
\begin{equation}
\left.\boldsymbol{R}_{\ell} \,\middle|\, \boldsymbol{R}_{\ell-1}\right.
\sim
    {\mathbb{N}(\boldsymbol{R}_{\ell-1},2\mathbb{I}\mathcal{D}\mathit{\Delta}{t}_{\ell})}
    +
    \boldsymbol{v}\mathit{\Delta}{t}_{\ell}.
\label{Equation:DBM}
\end{equation}
We restricted drift to the azimuthal plane $\boldsymbol{v}=v(\hat{\mathbf{x}} + \hat{\mathbf{y}})$ to keep particles in focus and drew a constant velocity from $v\sim\mathbb{U}_{(0,1]} \, (\mathrm{\upmu m} / s)$ to keep them in frame, enforcing the lower limit through a machine precision of $\varepsilon\approx10^{-16}$.
}}

\noindent \fbox{\parbox{\linewidth}{
\textbf{ATTM}: Particles undergoing ${\text{ATTM}}$ exhibit localized patches of $\text{BM}$, periodically redrawing power-law-distributed local diffusivity $\mathcal{D}_{\ell}$ after sojourn duration $\tau_{\ell}$\autocite{Manzo2014ATTM}. The generative model for such trajectories is
\begin{align}
\notag
    P(\mathcal{D}) 
&  \sim
    \mathcal{D}^{\varsigma-1}, \quad \quad \quad \quad \,\varsigma>0    
\\
\notag
    \mathcal{D}_{\ell}
& \sim
    P(\mathcal{D})
\\  \tau_{\ell}
\notag
& \sim
    \delta(\tau_{\ell} - \mathcal{D}_{\ell}^{-\gamma}), \quad \gamma \in (\varsigma,\varsigma+1)
\\ \notag
\mathit{\Delta}{t}_{\ell}
& =
    \tau_{\ell} - t_{\ell-1}
\\  
\left.\boldsymbol{R}_{\ell} \,\middle|\, \boldsymbol{R}_{\ell-1},\mathcal{D}_{\ell},\tau_{\ell}\right.
& \sim
    {\mathbb{N}(\boldsymbol{R}_{\ell-1},2\mathbb{I}\mathcal{D}_{\ell}\mathit{\Delta}{t}_{\ell})}.
\label{Equation:ATTM}
\end{align}
In this model, the anomalous exponent becomes $\alpha=\varsigma/\gamma$.
 }}

\noindent \fbox{\parbox{\linewidth}{
\textbf{CTRW}: In CTRW, a walker's spatial displacement $\mathit{\Delta}{\boldsymbol{R}}_{\ell}$ is decoupled from the power-law-distributed sojourn time $\tau_{\ell}$ for which it is transiently trapped\autocite{Montroll1965, Metzler2011CTRW}. The generative model for the trajectory of a particle evolving according to $\mathrm{CTRW}$ is
\begin{align}
    \notag
    P(\tau)
& \sim
    \tau^{-(\alpha+1)}, \quad \alpha \in (0, 1)
\\
    \notag
    \tau_{\ell}
& \sim
    P(\tau)
\\  \notag
    \mathit{\Delta}{\boldsymbol{R}}_{\ell}
& \sim
    {\mathbb{N}(\boldsymbol{0}, \, 2\mathbb{I}D\tau_{0})}
\\  \boldsymbol{R}_{\ell} 
& = 
    \boldsymbol{R}_{\ell-1} + \mathit{\Delta}{\boldsymbol{R}}_{\ell},
\label{Equation:CTRW}
\end{align}
where $D$ is the rescaled mobility parameter, and $\tau_{0}$ is arbitrary.
}}

\noindent \fbox{\parbox{\linewidth}{
\textbf{FBM}: $\mathrm{FBM}$ is a zero-mean Gaussian stochastic process with long-range temporal correlations. Its self-similarity is quantified by the Hurst index $\mathcal{H}\!\equiv\!\alpha/2\!\in\!(0,1)$ \autocite{Mandelbrot1968FBM}. This index governs the process' expectation value
\begin{equation}
    \mathbb{E}\!\left[  \left(  \boldsymbol{R}_{t}-\boldsymbol{R}_{0} \right)   \otimes
        \left(  \boldsymbol{R}_{t+\mathit{\Delta}{t}}-\boldsymbol{R}_{t}    \right)    \right]
=
    \mathbb{I} \! \left[  (t+\mathit{\Delta}{t})^{2\mathcal{H}}   -   t^{2\mathcal{H}}    -   \mathit{\Delta}{t}^{2\mathcal{H}}    \right]/2
\label{Equation:Mean(FBM)}
\end{equation}
covariance
\begin{equation}
        \langle{\boldsymbol{R}_{t}\boldsymbol{R}_{T}}\rangle
    =
        \mathbb{I}
        \mathcal{D}_{2\mathcal{H}}
        \left(
              t^{2\mathcal{H}}
            +T^{2\mathcal{H}}
            -\left|{t-T}\right|^{2\mathcal{H}}
        \right),
\label{Equation:Covariance(FBM)}
\end{equation}
and autocovariance function
\begin{equation}
        \gamma_\ell
    =
        \left(
                 \left|\ell+1\right|^{2\mathcal{H}}
            -2\left|\ell\right|^{2\mathcal{H}}
            +\left|{\ell-1}\right|^{2\mathcal{H}}
        \right)/2,
\label{Equation:Autocovariance(FBM)}
\end{equation}
where $\otimes$ denotes the outer product. Below, we specify the Davies-Harte circulant-embedding algorithm\autocite{DaviesHarte1987HurstEffect} for sampling $\mathrm{FBM}$ trajectories, but there exist alternate algorithms (\textit{e}.\textit{g}., the Hosking algorithm \autocite{Hosking1984FractionalDifferencing} and the Cholesky method \autocite{Asmussen2006FBM}. This algorithm simultaneously generates all $L$ increments by Fourier-diagonalizing fractional Gaussian noise's circulant autocovariance matrix $\mathbf{circ}(\{\gamma_{\ell}\})$, scaling white-noise by its eigenvalue square roots, and applying an inverse Fourier transform $(\mathfrak{F}^{-1})$ to recover increments in the time-domain. Denoting the complex normal distribution as $\mathbb{N}_{\mathbb{C}}$ and complex conjugation as $^\dagger$, the generative model can be written succinctly as 
\begin{align}
\notag
    \chi_{1,L}^{1:3}   & \sim    \mathbb{N}(0,1)
\\ \notag
    \chi_{2:L-1}^{1:3}    &\sim   \mathbb{N}_{\mathbb{C}}(0,1)
\\ \notag
    \chi_{2L-\ell}^{1:3}  &=  (\chi_{\ell}^{1:3})^{\dagger}
\\
    \boldsymbol{R}_{\ell}
&=
    \sum_{\ell^{\prime}=1}^{\ell}
    \mathfrak{R}    \!
    \left[
        \mathfrak{F}^{-1}   \!
            \left(\boldsymbol{\Lambda}
            \odot \boldsymbol{\chi}
            \right)
    \right],
\label{Equation:FBM}
\end{align}
where $\boldsymbol{\Lambda}\in\mathbb{R}^{2L\times3}$ broadcasts the square roots of the circulant matrix's eigenvalues $(\sqrt{\boldsymbol{\lambda}}\in\mathbb{R}^{2L})$ across three columns, $\boldsymbol{\chi}\in\mathbb{C}^{2L\times3}$ denotes the complex spectral noise matrix, and $\mathfrak{R}$ extracts the real part.
}}

\noindent \fbox{\parbox{\linewidth}{
\textbf{LW}: 
L\'{e}vy walks combine heavy-tailed jump-time statistics with a constant finite velocity $\boldsymbol{v}=v(\hat{\mathbf{x}}+\hat{\mathbf{y}}+\hat{\mathbf{z}})$, so each displacement spans the distance traveled in power-law-distributed jump time $\tau_{\ell}^{\prime}$\autocite{Klafter2015LevyWalk}. Hence, the generative model for particle trajectories is given by
\begin{align}
    \notag
    P(\tau^{\prime})%
& \sim
    {\tau^{\prime}}^{-(\varsigma+1)}, \quad \varsigma \in (0, 2)
\\
    \notag
    \tau_{\ell}^{\prime}
& \sim
    P(\tau^{\prime})
\\ \left. \boldsymbol{R}_{\ell} \middle| \boldsymbol{R}_{\ell-1},\tau_{\ell}^{\prime} \right.
& \sim
    \boldsymbol{R}_{\ell-1} \pm \boldsymbol{v}\tau_{\ell}^{\prime}.
\label{Equation:LW}
\end{align}
For $\varsigma\in(0,1)$, the particle undergoes ballistic diffusion because the anomalous exponent becomes $\alpha=2$; for $\varsigma\in(1,2)$, the particle exhibits superdiffusion from an anomalous exponent of $\alpha=3-\varsigma$. \textit{N}.\textit{B}.: in \cref{Equation:LW}, displacement directions are sampled isotropically from the unit sphere to ensure a uniform angular distribution.
}}

\noindent \fbox{\parbox{\linewidth}{
\textbf{SBM}: The Langevin formulation of $\mathrm{SBM}$\autocite{Metzler2014sBm}, $\dot{\boldsymbol{R}}(t)=\boldsymbol{\xi}(t)\sqrt{2\alpha \mathcal{D}_{\alpha} t^{\alpha-1}}$ for white Gaussian noise $\boldsymbol{\xi}(t)$, leads to independent Gaussian increments with isotropic covariance matrix $\boldsymbol{\Sigma}_{\ell}={2\mathbb{I}\mathcal{D}_{\alpha}\mathit{\Delta}{t}_{\ell}^{\alpha}}$. Thus, we generate particle trajectories through the following generative model
\begin{align}
    \notag
    \mathcal{D}_\alpha 
& = 
    \frac{1}{2\,\Gamma(1+\alpha)}
\\
    \left.  
    \boldsymbol{R}_{\ell}   
    \,  \middle|  \,    \boldsymbol{R}_{\ell-1} \right.
& \sim
    \mathbb{N}(\boldsymbol{R}_{\ell-1}, \, 
        {2\mathbb{I}\mathcal{D}_{\alpha}\mathit{\Delta}{t}_{\ell}^{\alpha}}),
\label{Equation:SBM}
\end{align}
where $\Gamma(\cdot)$ denotes the Gamma function.
}}

\subsubsection*{Generating Imaging Data}
\par To generate imaging data, we calculated the photoelectron load $\boldsymbol{u}$ from photons emitted from the fluorophore along the particle's trajectory $\boldsymbol{R}_{1:L}$ and from background illumination throughout the imaging plane using \cref{Equation:Expectation}. We then obtained synthetic measurements $\boldsymbol{w}$ after corrupting the photoelectron load with EMCCD measurement noise using \cref{Equation:Measurement}.

\subsection*{Evaluating the Likelihood's Probabilistic Weights}
\par Having generated particle positions evolving according to each motion model independently, we simulated fluorescence alongside static and dynamic localization errors using \cref{Equation:Expectation}. To then synthetically image particle trajectories, we degraded expectation frames through a stochastic measurement process with Gamma-distributed noise reminiscent of the post-multiplication electron count of an EMCCD camera\autocite{GammaEMCCD, Hirsch2013EMCCD, Ryan2021EMCCD}. Below, \cref{Table(Parameters)} provides a list of values assigned to emission, optical, and detector parameters chosen for consistency with typical values measured in widefield fluorescence SPT\autocite{Yildiz2003SingleFluorophoreImaging, Li2024StatisticalResolution, Zheng2014OrganicFluorophores, Grimm2015Fluorophores, Xiang2020DisplacementMapping}, as measured through the Andor iXon Ultra 888---a current fast megapixel, back-illuminated EMCCD---equipped with a Nikon CFI Plan Apochromat Lambda 1.45 NA ×100 oil objective and Olympus immersion oil. To demonstrate successful particle tracking against appreciable dynamic localization error arising from a detector with relatively poor temporal resolution, we considered the detector's full-frame exposure period of $\tau \approx 30 \, \mathrm{ms}$ and then cropped the imaging plane to a ROI spanning $32\times32\,\mathrm{pixels}$ and centered about the mean position of each particle's trajectory near the optical axis.
\begin{table}[H]
    \centering    
    \begin{tabular}{|ccc|}
 \hline
\multicolumn{3}{|c|}{\textbf{Emission Parameters}}
\\ \hline\hline
    \underline{Quantity}&\underline{Assigned Value}&\underline{Units}
\\
    \(\text{Fluorophore Count:}\)& \( M := 1 \quad \ \ \ \ \ \ \ \ \: \)&
    \\ \(\text{Ambient Photon Flux:}\)&$F:=10^{5}\,$\autocite{Li2024StatisticalResolution} $ \ \ \ \ \, $&\(\mathrm{1/\upmu m^{2}s}\)
    \\ \(\text{Photon Emission Rate:}\)&\( H:=10^{4}\,\)\autocite{Yildiz2003SingleFluorophoreImaging, Zheng2014OrganicFluorophores, Grimm2015Fluorophores}&\( \mathrm{1/s} \)
\\ \hline\hline
\multicolumn{3}{|c|}{\textbf{Optical Parameters}}
\\ \hline\hline
    \underline{Quantity}& \underline{Assigned Value}&\underline{Units}
\\
 \(\text{Numerical Aperture:}\)& \( N_\mathrm{A} := 1.45\, \)\autocite{Xiang2020DisplacementMapping} $ \quad \ \ $&
    \\ \(\text{Refractive Index:}\)& \( \mathfrak{n} := 1.515 \,\)\autocite{Fujiwara2020RefractiveIndex} $ \ $&
    \\ \(\text{Emission Wavelength:}\)& \( \lambda := 665 \ \ \, \,\)  $ \quad \, $&\( \mathrm{nm} \)
\\ \hline\hline
\multicolumn{3}{|c|}{\textbf{Detector Parameters  (EMCCD)}}
\\ \hline\hline
    \underline{Quantity}& \underline{Assigned Value}& \underline{Units}
\\ 
    \(\text{Gain:}\)& \( G := 350 \,\)\autocite{Hirsch2013EMCCD} $ \ \ \ \, $& 
    \\ \(\text{Conversion Factor:}\)& \( \varphi := 3.5 \,\)\autocite{Bernardes2018CCDs, Ryan2021EMCCD} $ \ $&$\mathrm{e^{-}/ADU}$
    \\ \(\text{Quantum Efficiency:}\)& \(\beta := 95\% \,\)\autocite{GammaEMCCD} $ \ \ \, $&
    \\
 \(\text{Imaging Area}^{\dagger}:\)& \( A := 32\times32 \,\) $ \ \, $&\( \mathrm{pixels} \)\\ \(\text{Frame Period}^{\dagger}:\)& \( t_\mathrm{F} := 33 \,\) $ \quad \quad \ \ \: $&\( \mathrm{ms} \)
    \\ \(\text{Exposure Period}^{\dagger}:\)& \( t_\mathrm{E} := 30 \,\) $ \quad \quad \ \ \: $&\( \mathrm{ms} \)
    \\ \(\text{Pixel Side Length}^{\ddagger}:\)& \( \mathit{\Delta}{s} := 133 \,\) $ \quad \quad \ \ \: $&\( \mathrm{n m} \)\\ \hline
    \end{tabular}
\caption{Reference values assigned for emission, optical, and detector parameters. 
The ambient photon flux contributing to background illumination is fixed at $F:=10^{5}\mathrm{/\upmu m^{2}s}$ to represent the mid-point of background ranges recently simulated in widefield fluorescence images\autocite{Li2024StatisticalResolution}; likewise, we set the photon emission rate contributing to the particle's fluorophore signal as 
$H:=10^{4} / \mathrm{s}$ to characterize inexpensive emitting particles [\textit{e}.\textit{g}., Cy3\autocite{Yildiz2003SingleFluorophoreImaging}, Cy5 without protective agents\autocite{Zheng2014OrganicFluorophores}, tetramethylrhodamine (TMR)\autocite{Grimm2015Fluorophores}, or JF549\autocite{Grimm2015Fluorophores}].
Some assigned parameters reference the Andor iXon Ultra 888 EMCCD, which has temporal resolutions 
between $26\,\mathrm{fps}$ for $1024\times1024$ pixels and $1319\,\mathrm{fps}$ for $64\times64$ pixels in the optically-centered crop mode suggested for widefield fluorescence microscopy, respectively. 
Through an objective lens with optical magnification $m=100$, the detector's pixel side length becomes $\mathit{\Delta}{s}/m=130\,\mathrm{nm}$ from $\mathit{\Delta}{s}=13\,\mathrm{\upmu m}$.
For this detector to provide a quantum efficiency (QE) of $\beta\approx95\%$ at room-temperature $T \approx 20\,{}^{\circ}\mathrm{C}$, we set our emission wavelength
to $\lambda=655 \, \mathrm{nm}$ despite it's peak QE $(\beta>95\%)$ existing between $\lambda\in(525,600)\,\mathrm{nm}$.
Furthermore, a large electron-multiplying gain $G>210$ was selected such that \cref{Equation:Measurement} models the measurement process accurately without Gaussian readout noise\autocite{Hirsch2013EMCCD}.
\\ ${}^{\dagger}$: The given frame period approximates that of the full-frame iXon Ultra 888, but we cropped the imaging plane to an ROI spanning $32\times32$ pixels.
\\ ${}^{\ddagger}$: The effective pixel side length is given with optical magnification.
}
\label{Table(Parameters)}
\end{table}

\subsection*{Motion Model Classification}
\par To evaluate the present experimental validity of methods devised to decode anomalous diffusion\autocite{Manzo2021AnDiChallenge}, we set out to determine whether or not pure diffusion could be discerned from anomalous diffusion using existing, commonly used tools featured in the AnDi Challenge\autocite{Manzo2021AnDiChallenge}: \textit{AnomDiffDB}\autocite{TeamJ2018Networks, TeamJ2019DeepLearning} and \textit{CONDOR}\autocite{CONDOR}. First, we gave motion model classification a complete advantage by inputting noiseless ground truth $\mathrm{BM}$ trajectories of diffusivities $\mathcal{D}:=\{1,5,10\} \, \mathrm{\upmu m^{2}/s}$. We then investigated how static and dynamic measurement noise biases motion model classifications by inputting trajectories inferred from synthetic observations using both \textit{TrackMate}\autocite{Trackmate, Trackmate7}, a conventional modular tracking algorithm, and \textit{BNP-Track}\autocite{Xu2024}, which features a joint posterior probability distribution associated with our likelihood \cref{Equation:Likelihood(General)}.

\section*{Discussion}
\par Anomalous diffusion has been ubiquitously invoked to model transport phenomena across heterogeneous and crowded environments, often featuring complex energy landscapes\autocite{Wong2004MicrostructureDynamics, Metzler2009AnalysisSPT, Metzler2014AnDiModels}. In particular, motion models of anomalous diffusion have been used to model drug and gene delivery through mucus layers\autocite{Metzler2019MucinHydrogels}, time-dependent temperatures\autocite{Metzler2016UnderdampedSBM, Metzler2016TimeFluctuating}, fluorophore photobleaching\autocite{Metzler2022sBmVERSUSfBm}, and diffusion through viscoelastic media (\textit{e.g.}, tissues)\autocite{Metzler2011CTRW, Metzler2013PowerLawRelaxation, Metzler2022RobustCriterion}.

Despite the widespread use of anomalous diffusion, motion model classification is a difficult problem, especially starting from widefield fluorescence data. Typical acquisition rates of conventional EMCCD\autocite{Kannan2006EMCCD} and sCMOS\autocite{Mandracchia2020sCMOS} cameras (\qtyrange{e2}{e3}{\fps}) yield dynamic localization errors that obscure the precise position of a particle, further introducing blurring artefacts even in the presence of modest diffusivities ($\mathcal{D}$\qty{>1}{\micro\m^2/\s}).

\par Due of these limitations, we find that diffusive trajectories can be accurately tracked by assuming the $\mathrm{BM}$ model regardless of the particle's underlying motion model. This is because the emission and measurement processes --- and not the dynamics --- primarily dictate our ability to track particles. To make this point clear, our likelihood formulation decouples contributions from the emission and motion models, and numerical evaluation shows that the emission term contributes over $99\%$ of the log likelihood across typical conditions.
\par Even when varying simulation parameters, the magnitude of the emission model remains at roughly $\ln \mathbb{P}\left(\text{Data} \,\middle|\, \text{Position}\right)\approx-10^5$, while that of the motion model is on the order of $\ln \mathbb{P}\left(\text{Motion}\right)\approx+10^3$ suggesting that the trajectories recovered are determined almost entirely by the photon emission and detection processes, not by the particle's physical motion. Thus, motion models play only a marginal role in trajectory inference.
These findings underscore the importance of accounting for realistic sources of noise through accurate emission models in data analysis and avoid reliance on post-processed trajectories, rather than raw imaging data, which may obscure the limited influence that motion models have on measurements.

\par Moreover, our analysis revealed a critical oversight in many existing software tools developed for anomalous diffusion analysis\autocite{AnDiRepository, Manzo2021AnDiChallenge}: they do not explicitly consider normal diffusion (\textit{i.e.}, $\mathrm{BM}$) as a candidate motion model. Specifically, $12/15 \ (80\%)$ competitors never confirmed whether input trajectories deviated from $\mathrm{BM}$\autocite{AnomalousUnicorns1992, AnomalousUnicorns2022, Krog2018fBmInference, LwInference, Gratin, DeepSPT2016DeepResidualLearning, DeepSPT2016ScalableBoosting, RANDI, WadNET, TeamH, ELM, TeamK, CONDOR, eRNN, TeamN2018, TeamN2019}. While many motion models of anomalous diffusion converge to $\mathrm{BM}$ at the Brownian limit ($\alpha=1$), \cref{Table(AnomalousInference)} highlights just how often pure $\mathrm{BM}$ is classified as anomalous. Further details highlighting how anomalous diffusion simulated at the Brownian limit was never identified as pure $\mathrm{BM}$ is relegated to the \textbf{Supplementary Data}. Furthermore, while many methods included in the first challenge could not analyze three-dimensional trajectories\autocite{AnomalousUnicorns1992, AnomalousUnicorns2022, Gratin, DeepSPT2016DeepResidualLearning, DeepSPT2016ScalableBoosting, TeamH, ELM, TeamJ2018Networks, TeamJ2019DeepLearning, TeamK}, importantly, none could make use of the real-world observations (\textit{i.e.}, image stacks) that we found contributing the vast amount of information necessary to track particle positions across motion models. These concerns remain despite the \textit{ad hoc} imposition of normally-distributed static localization error in the AnDi Challenge\autocite{Manzo2021AnDiChallenge}, as this purely random noise fails to approximate proper emission models\autocite{Ayush2023, Presse2024FluorescenceMicroscopy} and fails to address dynamic localization errors that arise at even modest diffusivities mentioned earlier ($\mathcal{D}$\qty{>1}{\micro\m^2/\s}).

\par Indeed, as we were able to ascribe $\approx\!99\%$ of the weight of the log likelihood in widefield SPT to the emission model rather than the particle's presumed dynamics, we make two suggestions for the purposes of motion model classification and parameter inference: 1) motion model inference must be approached with considerable caution --- few trajectories, short tracks, and low photon budgets typical of tracking experiments in widefield exacerbate the risk of over-interpreting noise as dynamical features; 2) motion model classification must start from the raw data (\textit{i}.\textit{e}., image stacks) where most of the information lies, not post-processed trajectories. 

\section*{Conclusion}
\par We find that trajectories of particles evolving according to anomalous diffusion models can be reliably inferred through Brownian (\textit{i.e.}, statistically independent, identically distributed, stationary Gaussian) transitions primarily due to the overwhelming contribution of the emission portion of the likelihood. As such, we can reliably track particles irrespective of which motion model generated the data. Along these lines, it may be more difficult to learn motion models in SNR regimes associated with diffraction-limited particle tracking at emission rates characteristic of endogenously expressed fluorescent labels\autocite{Coelho2013InVivo} or synthetic dyes $(H\!\!\approx\!\!10^{5}/\mathrm{s})$\autocite{Zheng2014OrganicFluorophores} versus the photon emission rate of quantum dots (QDs) $H\in[10^{6},10^{8}]/\mathrm{s}$\autocite{Tamariz2020QuantumDot, Abudayyeh2021QuantumDot}.

\par This work leaves to wonder whether existing analysis tools\autocite{AnomalousUnicorns1992, AnomalousUnicorns2022, Krog2018fBmInference, LwInference, Metzler2022sBmVERSUSfBm, Gratin, DeepSPT2016DeepResidualLearning, DeepSPT2016ScalableBoosting, RANDI, TeamF, WadNET, TeamH, ELM, TeamJ2018Networks, TeamJ2019DeepLearning, TeamK, CONDOR, eRNN, TeamO2019, TeamO2020, TeamO2020Impact, Szarek2021Neural, Das2009HMM, Manzo2021AnDiChallenge, NoBIAS, RANDI, MunozGil2021Unsupervised, Metzler2023ML, Qu2024Semantic} would benefit from avoiding the analysis of post-processed trajectories to learn motion models. Nonetheless, the first anomalous diffusion challenge\autocite{Manzo2021AnDiChallenge} had all $15$ competitors start from pre-localized data (post-processed trajectories): seven used post-processed as input\autocite{AnomalousUnicorns1992, AnomalousUnicorns2022, Krog2018fBmInference, LwInference, Metzler2022sBmVERSUSfBm, RANDI, WadNET, TeamH, TeamJ2018Networks, TeamJ2019DeepLearning, eRNN}, six used input features\autocite{TeamF, ELM, TeamK, CONDOR, TeamN2018, TeamN2019, TeamO2020, TeamO2020Impact}, and the remaining two took post-processed trajectories combined with features\autocite{Gratin, DeepSPT2016DeepResidualLearning, DeepSPT2016ScalableBoosting}. We believe that trajectory classification remains possible, though perhaps by working directly with raw data and working with new generations of dyes such as $\text{PF555}$, shown to provide a remarkably photostability and bright signal without the multivalency or bulkiness of QDs\autocite{Kim2025PhotostableDye}.

\section*{Acknowledgments}
\par We would like to thank Ayush Saurabh, Pedro Pessoa, Max Schweiger, and Ioannis Sgouralis for interesting discussions. SP acknowledges support from the NIH (R35GM148237), ARO (W911NF-23-1-0304), and NSF (Grant No. 2310610).

\section*{Code \& Data Availability}
\par All software \autocite{WideFieldMotionModels} developed for simulation, inference, and the analysis thereof is available on GitHub through an MIT license. A link to all data \autocite{Zenodo} collected throughout this study is available within the GitHub repository.

\section*{Author Contributions}
Each author contributed meaningfully to the study's conception, execution, and manuscript writing. All authors reviewed and approved the final version. SP oversaw all aspects of the project.

\section*{Competing Interests}
The authors declare no competing interests.

\printbibliography[heading=bibintoc,title={References (Main)}]
\end{refsection}

\clearpage
\section*{Supplementary Information}
\begin{refsection}

    \crefformat{equation}{#2SI  Equation~#1#3}
    \crefformat{figure}{#2SI  Figure~#1#3}
    \crefformat{table}{#2SI  Table~#1#3}
    \crefrangeformat{figure}{SI Figures #3#1#4--#5#2#6}
    

\def\ppi{{\usefont{U}{bboldx}{m}{n}\char25}} 
\def\PPi{{\usefont{U}{bboldx}{m}{n}\char5 }} 
\def\00{{\usefont{U}{bboldx}{m}{n}\char48 }} 

\renewcommand{\figurename}{SI Figure}

\tableofcontents 

\section*{Part I: Burn In Removal}
\addcontentsline{toc}{section}{Part I: Burn In Removal}
\par Markov Chain Monte Carlo (MCMC) samplers are often started from an intentionally over-dispersed initial guess --- a point in the parameter space that is unlikely to coincide with the data-driven mode. 
Accordingly, the early search resembles a biased exploration, and this burn in (\textit{i}.\textit{e}., initialization bias) should be removed to ensure that posterior summaries reflect equilibrium sampling. In \cref{FigureSI:BurnInRemoval} below, we show inferred values of diffusivity $\mathcal{D}$ from $I=2000$ MCMC samples. Since the first $\approx50\%$ of samples are non-stationary, we keep only the second half of samples $i\in[1,2]\cdot10^{3}$; consequently, the mean over kept iterations adequately represents the high posterior density and approximates ground truth.
\begin{figure}[H]
    \centering
    \includegraphics[width=0.5\linewidth]{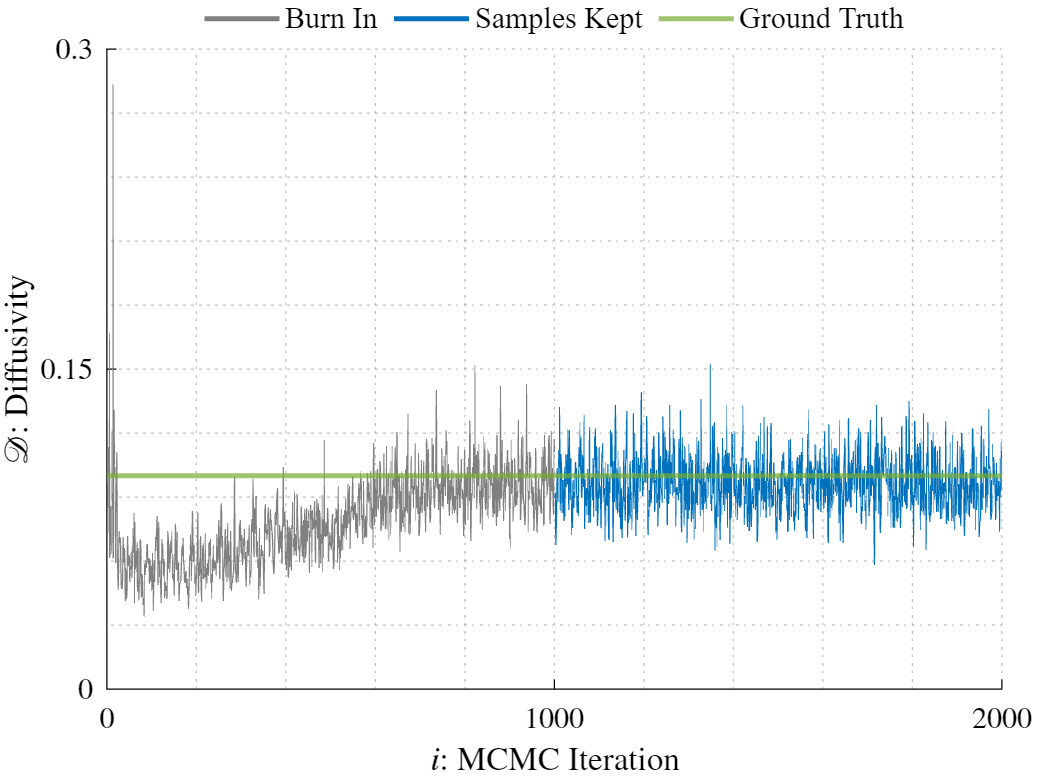}
    \caption{Inferred diffusivity $\mathcal{D}\,(\mathrm{\upmu m^{2}/s})$ for a single Brownian walker from an MCMC chain of $I=2000$ iterations. Samples from the initial $999$ iterations (gray) constitute the burn-in phase; they display a clear upward drift as the chain moves from its over-dispersed starting point toward the high-posterior-density region. The remaining $1000$ iterations (blue) fluctuate symmetrically around the ground truth $\mathcal{D}$ (green), indicating that the chain has reached stationarity. Discarding the burn-in therefore removes initialization bias while preserving samples that faithfully represent the target posterior distribution.}
    \label{FigureSI:BurnInRemoval}
\end{figure}

\section*{Part II: Alternate Motion Models} 
\addcontentsline{toc}{section}{Part II: Alternate Motion Models for Anomalous Diffusion}
\par Here, we present additional motion models that can be incorporated into our likelihood. 
In what follows, we describe the defining features and probabilistic formulations for each candidate model. Because the formulations for $\mathrm{FBM}$, $\mathrm{LW}$, and $\mathrm{SBM}$ are intended for equally spaced lag times, interpolation is not advised; therefore, $L$ should be defined as the number of observed frames ($N$ in the main text).

\subsection*{DBM Motion Model}
\addcontentsline{toc}{subsection}{Directed Brownian Motion}
\par Directed Brownian motion $(\mathrm{DBM})$ follows the same stochastic framework as Brownian motion $(\mathrm{BM})$ model (Equation 3 in the main text) --- namely, a sequence of Brownian (\textit{i.e.}, statistically independent, stationary, Gaussian) transitions. However, this motion model allows for a constant drift speed $v$ that biases each displacement in a fixed direction $\hat{\boldsymbol{v}}$. Accordingly, the evolution of the particle's position is governed by both stochastic diffusion and deterministic drift, with the resulting motion model given by
\begin{equation}
\mathbb{P}\!\left(\mathit{\Delta}\boldsymbol{R}_{1:L}\right)
=
\exp\!
\left[
    {-
        \sum\limits_{\ell = 1}^{L}
            \frac{\left|\mathit{\Delta}\boldsymbol{R}_{\ell} - \boldsymbol{v}\mathit{\Delta}{t}_{\ell}\right|^{2}}
            {4\mathcal{D}\mathit{\Delta}t_{n}^{k}}}
            \right]
        \prod_{\ell=2}^{L}
        \left(4\uppi\mathcal{D}\mathit{\Delta}{t}_{\ell}\right)^{-3/2},
        \label{Equation:MotionModel(DBM)}
\end{equation}
where $\boldsymbol{v}\equiv v\hat{\boldsymbol{v}}$ is the particle's velocity vector, $\mathit{\Delta}{t}_{\ell}\equiv t_{\ell}-t_{\ell-1}$ is the lag-time between particle positions spanning a Euclidean distance of $\left|\mathit{\Delta}\boldsymbol{R}_{\ell}-\boldsymbol{v}\mathit{\Delta}{t}_{\ell}\right|$.

\subsection*{FBM Motion Model}    
\addcontentsline{toc}{subsection}{Fractional Brownian Motion}
\par We generalize the one-dimensional motion model\autocite{Krog2018fBmInference, Metzler2022sBmVERSUSfBm} for $\mathrm{FBM}$ to three-dimensional space as
\begin{equation}
    \left.{\mathbb{P}}(\mathit{\Delta} \boldsymbol{R}_{1:L}
    \right)
=
    \left[(2\uppi)^{L}
    \left| \boldsymbol{\Sigma}_{L} \right|\right]^{-\frac{3}{2}}
        \exp\!\left[
        -\frac{1}{2} 
        \sum_{\ell=1}^{L}
          \mathit{\Delta} \boldsymbol{R}_{\ell}^{\top}
          \left( \boldsymbol{\Sigma}_{L}^{-1} 
          \otimes \mathbb{I} \right) \,
      \mathit{\Delta} \boldsymbol{R}_{\ell}
    \right],
\label{Equation:MotionModel(FBM)}
\end{equation}
where $\boldsymbol{\Sigma}_{L}$ is the covariance matrix with elements $(\mathit{\Sigma}_{L})_{\ell}^{{\ell^\prime}} = \gamma(\ell-\ell^{\prime})$ defined through the autocovariance function
\begin{equation}
    \gamma(\ell)
=
    \mathcal{D}_{\mathcal{H}}t_{E}^{2\mathcal{H}}
    \left[
         \left|\ell+1\right|^{2\mathcal{H}}
       +\left|\ell-1\right|^{2\mathcal{H}}
        -2\left|\ell\right|^{2\mathcal{H}}
    \right],
\label{EquationSI:Autocovariance(FBM)}
\end{equation}
and $\otimes$ denotes the Kronecker product. Because inversion of $\boldsymbol{\Sigma}_{L}$ is computationally expensive, we explore a second means by which \cref{Equation:MotionModel(FBM)} can be calculated: the likelihood can be rewritten using the chain rule of conditional probabilities:
\begin{equation}
    \mathbb{P}\!\left(\mathit{\Delta}{\boldsymbol{R}_{1:L}} 
    \right)
=
    (2\uppi)^{-\frac{3L}{2}}
    \prod_{\ell=1}^{L}
    \frac{1}{{\varsigma_{\ell}^{3}}}
    \mathrm{exp\left[-
    \frac{(\mathit{\Delta{\boldsymbol{R}}_{\ell}}-\mathit{\Delta}{\boldsymbol{\mu}_{\ell}})^{2}}
        {2\varsigma_{\ell}^{2}}
                                            \right]}
,
\label{Equation:MotionModel(FBM2)}
\end{equation}
where the mean $\boldsymbol{\mu}$ and standard deviation $\boldsymbol{\varsigma}$ are iterated using the Durbin-Levinson algorithm~\cite{Brockwell1991TimeSeries, Krog2018fBmInference}.  For iteration of the mean and standard deviation, we initialize $\mathit{\Delta} \boldsymbol{\mu}_1 = \boldsymbol{0}$ and $\sigma_1^2 = \gamma(0)$ before recursively iterating successive values through
\begin{equation}
\begin{cases}
\mathit{\Delta} \boldsymbol{\mu}_{\ell+1} & = \sum\limits_{\ell^{\prime}=1}^{\ell} \phi_{\ell}^{\ell^{\prime}}  \,
\mathit{\Delta} \boldsymbol{R}_{\ell+1-\ell^{\prime}},
\\
\varsigma_{\ell+1}^2 & = \varsigma_{\ell}^2 \left[ 1 - (\phi_{\ell}^{\ell})^2 \right],
\end{cases}
\end{equation}
where the coefficients $\phi_{\ell}^{\ell^{\prime}}$ are obtained using \cref{EquationSI:Autocovariance(FBM)} as
\begin{equation}
\begin{cases}
\phi_{1}^{1} = \gamma(1)/\gamma(0)
\\
\phi_{\ell}^{\ell} = {\sigma_{\ell}^{-2}} \, \gamma(\ell) - \sum\limits_{\ell^{\prime}=1}^{\ell-1} \gamma(\ell-\ell^{\prime}) \, \phi_{\ell-1}^{\ell^{\prime}},
\\
\phi_{\ell}^{\ell^{\prime}} = \phi_{\ell-1}^{\ell^{\prime}} - \phi_{\ell-1}^{\ell-\ell^{\prime}} \phi_{\ell}^{\ell}, \quad  1 \leq \ell^{\prime} < \ell.
\end{cases}
\end{equation}
\par \textbf{Prior Distributions:} Since the Hurst parameter exists along the interval $\mathcal{H}\in(0,1)$, modeling it as a uniformly distributed hyperparameter gives its distribution as $\mathbb{P}(\mathcal{H})=1$. To then assign prior knowledge to the anomalous diffusivity, we prescribe a Jeffreys' prior\autocite{Presse2023} on the standard deviation for a single step $\sigma_{\mathcal{H}}$:
\begin{equation}
    \mathbb{P}(\mathcal{\sigma_{\mathcal{H}}})
=
\begin{cases}
    \left(\sigma_{\mathcal{H}} 
    \ln{\dfrac{\overline{\mathcal{\sigma}}_{\mathcal{H}}}{\underline{\mathcal{\sigma}}_{\mathcal{H}}}}
    \right)^{-1}
&, \quad
\mathcal{\sigma_{\mathcal{H}}}\in[\underline{\mathcal{\sigma}}_{\mathcal{H}}
        ,\overline{\mathcal{\sigma}}_{\mathcal{H}}]
\\
0&,\quad 
    \mathcal{\sigma_{\mathcal{H}}} 
    \;\slashed{\in}\;
    [\underline{\mathcal{\sigma}}_{\mathcal{H}}
        ,\overline{\mathcal{\sigma}}_{\mathcal{H}}].
\end{cases}
\label{Equation:Prior(FBM,StandardDeviation)}
\end{equation}
Doing so allows us to define the generalized diffusivity\autocite{Krog2018fBmInference} as
\begin{equation}
    \mathcal{D}_{\mathcal{H}}
=
    \frac{\sigma_{\mathcal{H}}^{2}}
                {2t_{E}^{2\mathcal{H}}}.
\label{Equation:Diffusivity(FBM)}
\end{equation}

\subsection*{LW Motion Model}      
\addcontentsline{toc}{subsection}{L\'{e}vy Walks}
\par Although there exists a one-dimensional motion model for $\mathrm{LW}$\autocite{Krog2018fBmInference}, it was parameterized upon an \textit{ad hoc} global step deviation. Accordingly, we must first re-parameterize the model to account for realistic static and dynamic localization errors. From known localization formulae\autocite{Thomson2002PreciseLocalization}, we consider the static and dynamic localization error to manifest as
\begin{equation}
    \varsigma_{\ell}^{p}
=
\sqrt{
    \frac{\sigma_{xy}^{2}+\mathit{\Delta}{s}^{2}/12}
                {\widetilde{C}_{\ell}^{p}}
+   \frac{8\uppi\sigma_{xy}^{4}(B_{\ell}^{p})^{2}}
        {(\widetilde{C}_{\ell}^{p})^{2} \mathit{\Delta}{s}^{2}}
+    \frac{\sigma_{\text{read}}^{2}}{(\widetilde{C}_{\ell}^{p})^{2}}
+     \frac{v^{2}\mathit{\Delta}{t}_{\ell}^{2}}
                   {12}
},
\label{Equation:LocalizationError}
\end{equation}
where $\widetilde{C}_{\ell}^{p}\equiv \beta (G/\varphi) u_{\ell}^{p}$ represents the photons counted at the $p$\textsuperscript{th} pixel in the $\ell$\textsuperscript{th} frame from photoelectron load $u_{\ell}^{p}$ apportioned by quantum efficiency $\beta$, dimensionless gain $G$, and the calibration parameter $\varphi$; $\sigma_{xy}$ is the PSF's axial width, $\mathit{\Delta}{s}$ is the pixel side length, $(B_{\ell}^{p})^{2}=\beta (G/\varphi)^{2}FA^{p}\mathit{\Delta}{t}_{\ell}+\sigma_{\text{read}}^2$ is the constant background noise, and $v^{2}\mathit{\Delta}{t}_{\ell}^{2}/12$ is a L\'{e}vy walker's dynamic localization error for constant speed $v$ over the exposure time $\mathit{\Delta}{t}_{\ell}\equiv t_{E}$. If we constrain our investigation to EMCCDs operated at high gain $G$, then background noise is dominated by Poissonian photon statistics such that the readout noise can be neglected: $B_{\ell}^{2}\approx G^{2}FAt_{E}$. Having defined the standard deviation of localization in the $\ell$\textsuperscript{th} frame as $\varsigma_{\ell}$ through \cref{Equation:LocalizationError}, we now prescribe the $\mathrm{LW}$ motion model for experimental imaging data as
\begin{equation}
    \mathbb{P}\!\left(
    \mathit{\Delta}{\boldsymbol{R}_{1:L}}
    \right)
=
    \prod_{\ell=1}^{L}
    \prod_{p=1}^{P}
    \left[4\uppi(\varsigma_{\ell}^{p})^{2}\right]^{-\frac{3}{2}}
    \exp\!
        \left(-
\dfrac{
\left|
\mathit{\Delta}\boldsymbol{R}_{\ell} - vt_{E}{\hat{\boldsymbol{v}}}_{\ell}
\right|^{2}
}
{4(\varsigma_{\ell}^{p})^{2}}
                    \right)
\end{equation}
for jump directions sampled on the the unit sphere $\hat{\boldsymbol{v}}_{\ell}\sim\mathbb{U}[\mathbb{S^{2}}]$.

\par \textbf{Prior Distributions}: Since $\mathrm{LW}$ only models superdiffusion and ballistic diffusion, the anomalous exponent is constrained to exist along the interval $\alpha\in(1,2]$; modeling $\alpha$ as a uniformly distributed hyperparameter, then, gives $\mathbb{P}(\alpha) = 1$. Given that random jumps of length $vt_{E}$ with constant velocity $v\sim\mathbb{U}_{[0,10]}$ were sampled on the the unit sphere $\hat{\boldsymbol{v}}\sim\mathbb{U}[\mathbb{S^{2}}]$, we need only further address the radial direction associated with each jump. That is, we adopt the radial vector described by length $r=vt_{E}$, azimuth $\phi\sim\mathbb{U}_{[0,2\uppi]}$, and zenith $\theta\sim\mathbb{U}_{[0,\uppi]}$. Accordingly, our joint prior is written
$
\mathbb{P}(\alpha)\,\mathbb{P}(v)\,\mathbb{P}(\theta)\,\mathbb{P}(\phi) = \frac{1}{20\uppi^{2}}$.

\subsection*{SBM Motion Model}    
\addcontentsline{toc}{subsection}{Scaled Brownian Motion}
\par We generalize $\mathrm{SBM}$'s motion model\autocite{Metzler2022sBmVERSUSfBm} to 
\begin{equation}
    \mathbb{P}\!\left(\mathit{\Delta}\boldsymbol{R}_{1:L}
    \right)
=
(2\uppi)^{-\frac{3L}{2}}
    \prod_{\ell=1}^{L}
        \frac{1}{\varsigma_{\ell}^{3}}
    \exp\left(-\frac{\left|\mathit{\Delta}{\boldsymbol{R}_{\ell}}\right|^{2}}
    {2\varsigma_{\ell}^{2}}\right),
\end{equation}
where $\tau$ is the system's aging prior to $t=0$, and the time-dependent standard deviation for a single step is 
\begin{equation}
    \varsigma_{\ell}^{2}
=
    2\mathcal{D}_{\alpha}
    t_{E}^{\alpha}
    \left[
        \left(\ell+\frac{\tau}{t_{E}}\right)^{\alpha}
        -
        \left(\ell-1+\frac{\tau}{t_{E}}\right)^{\alpha}
    \right].
    \label{Equation:StandardDeviation(SBM)}
    \end{equation}
\par \textbf{Prior Distributions}: Since $\mathrm{SBM}$ describes subdiffusion, $\mathrm{BM}$, superdiffusion, and ballistic diffusion, we model the anomalous exponent as a uniformly distributed hyperparameter along the interval $\alpha\in(0,2]$, which returns its distribution as $\mathbb{P}(\alpha)=1/2$. Following the one-dimensional formulation\autocite{Metzler2022sBmVERSUSfBm}, we prescribe a standard normal distribution $\mathbb{N}(0,1)$ on $\log_{10}{\varsigma_{1}}$ such that 
$\mathbb{P}\left(\varsigma_{1}\right) = (\varsigma_1 \ln{10} \sqrt{2\uppi})^{-1}\exp\left[ -\left(\log_{10} \varsigma_{1}\right)^{2}/2\right]$.

\section*{Part III: Alternate Emission Models}
\addcontentsline{toc}{section}{Part III: Alternate Emission Models}
        \subsection*{EMCCD Poisson-Gamma-Normal (PGN) Noise Model}   
\addcontentsline{toc}{subsection}{Emission Model for Poisson-Gamma-Normal (PGN) Noise Model of EMCCD Detectors}
\subsubsection*{Step 1: Photon Arrival \& Photoelectron Genesis}
\par Light captured by a detector comes from both signal and noise. The incident photon flux per pixel per frame, given a photon emission rate $H$ contributing to fluorophore signal in an environment of ambient photon flux $F$ contributing to background illumination, is 
\begin{equation}
    \mathit{\mathit{\Phi}}_{n}^{p} 
\equiv 
    \left[F+H\,\mathrm{PSF}(x^p,y^p\,;\,\boldsymbol{R}_{n})\right] \, \mathrm{d}A^{p} \, \mathrm{d}t_{n},
\label{Equation:Phi}
\end{equation}
where  $\mathrm{d}t_{\ell} \equiv t_{E}$ is merely the exposure period, and $\mathrm{d}A^{p}$ is the area of the $p$\textsuperscript{th} pixel. Upon capturing this light, $\mathit{\Phi}_{n}^{p}$ gets apportioned by the detector's quantum efficiency $\beta$ characterizing its ability to generate a photoelectron per incident photon. Additionally, thermal noise manifests as ``dark" photoelectron counts $\slashed{C}_{n}^{p}\equiv\dot{\slashed{C}}_{n}^{p}\,\mathrm{d}A^{p}\mathrm{d}t_{n}$ arising from the dark current $\dot{\slashed{C}}^{p}$ over area $\mathrm{d}A^{p}$ in duration $t_{E}$. Furthermore, the clock-induced charge (CIC) $c$ spuriously generates electrons, adding an exposure-independent Poisson offset to pixels. Since any source of light consists of discrete quanta, the detector's pre-amplification photoelectron load $C_{n}^{p}$ in frame $n$ at pixel $p$ is Poisson distributed:
\begin{equation}
    \mathbb{P}\!\left(C_{n}^{p} \,\middle|\, \mathit{\Phi}_{n}^{p}\right)
= 
    \frac{(\beta \mathit{\Phi}_{n}^{p} + \slashed{\dot{C}}^{p}t_{E} + c)^{C_{n}^{p}}}{C_{n}^{p}!} 
    \exp\!\left[ -(\beta \mathit{\Phi}_{n}^{p} + \slashed{\dot{C}}^{p}t_{E} + c) \right].
\label{Equation:Poisson(PGN)}
\end{equation}

\subsubsection*{Step 2: Stochastic Multiplication in EM Register}
\par In an EMCCD camera, photoelectrons comprising the pre-load $C_{n}^{p}$ traverse an electron-multiplying (EM) gain register at random, yielding a post-amplification photoelectron load ${C^{\prime}}_{n}^{p}$ that follows Tubb's distribution 
\begin{equation}
    \mathbb{P}\!\left({C^{\prime}}_{n}^{p}  \, \middle| \, C_{n}^{p}
    \right)    
=
    \dfrac{({C^{\prime}}_{n}^{p}-C_{n}^{p}+1)^{C_{n}^{p}-1}}{(C_{n}^{p}-1)!(G+1/C_{n}^{p}-1)^{C_{n}^{p}}}
    \exp{\left(-
            \dfrac{{C^{\prime}}_{n}^{p}-C_{n}^{p}+1}
                          {G+1/C_{n}^{p}-1}
          \right)},
\label{PGN(Multiplication)}
\end{equation}
where $G$ is the gain and $0<C_{n}^{p}\le {C^{\prime}}_{n}^{p}$. This stage introduces the multiplicative noise with an exponential tail that is characteristic of EMCCDs\autocite{Hirsch2013EMCCD}.

\subsubsection*{Step 3: Analogue Read Out}
\par During read-out, the amplified photoelectron load ${C^{\prime}}_{n}^{p}$ gets converted to voltage before being digitized as an analog-to-digital unit (ADU) count $w_{n}^{p}$. This read-out process introduces a zero-mean Gaussian noise with standard deviation $\sigma$. Thus, the probability distribution of raw measurements $w_{n}^{p}$ is read
\begin{equation}
    \mathbb{P}\!\left(w_{n}^{p} 
    \,\middle|\, {C^{\prime}}_{n}^{p}
    )
    \right. 
=
    (2\uppi)^{-\frac{1}{2}}
    \frac{1}{\sigma^{p}} 
    \exp\!\left[ -\frac{(w _{n}^{p}- G^{p}{C^{\prime}}_{n}^{p})^{2}}{2(\sigma^{p})^{2}} \right].
\end{equation}
\subsubsection*{The PGN Emission Model}
\par Propagating uncertainties from photon statistics, stochastic multiplication, and the final output entails convolving the Poisson, EM-gain, and read-out stages. This Poisson-Gamma-Normal ($\mathrm{PGN}$) noise model reads
\begin{equation}
    \mathbb{P}\!\left(w_{1:N}^{1:P}  \, \middle| 
    C_{1:N}^{p},{C^{\prime}}_{1:N}^{1:P}\right.)    
=
    \prod_{n=1}^{N}
    \prod_{p=1}^{P}
    \left[
        \sum_{C_{n}^{p}=0}^{\infty}
        \sum_{{C^{\prime}}_{n}^{p}=0}^{\infty}
    \mathbb{P}\!\left(C_{n}^{p} \,\middle|\, \mathit{\Phi}_{n}^{p}\right)
    \,
    \mathbb{P}\!\left({C^{\prime}}_{n}^{p}  \, \middle| \, C_{n}^{p} \right.)    
  \,
    \mathbb{P}\!\left(w_{n}^{p} 
    \,\middle|\, {C^{\prime}}_{n}^{p})
    \right.
\right]
.
\label{Equation:EmissionModel(PGN)}
\end{equation}
\cref{Equation:EmissionModel(PGN)} above replaces Equation 6 in the main text for any values of $w_{n}^{p}$, $G$, and $\sigma^{p}$; the sums can be truncated at sufficiently high counts for a satisfactory approximation.

        \subsection*{sCMOS Detector Architecture}                                            
\addcontentsline{toc}{subsection}{Emission Model for sCMOS Detectors}
\subsubsection*{Step 1: Photon Arrival \& Photoelectron Genesis}
\par Once more, we define the incident photon flux per pixel per frame from \cref{Equation:Phi} as 
\[ \mathit{\mathit{\Phi}}_{n}^{p} \equiv \left[F+H\,\mathrm{PSF}(x^p,y^p\,;\,\boldsymbol{R}_{n})\right] \, \mathrm{d}A^{p} \, \mathrm{d}t_{n},
\]
where $H$ is the photon emission rate contributing to fluorophore signal in an environment of ambient photon flux $F$ contributing to background illumination; $\mathrm{d}A^{p}$ is the area of the $p$\textsuperscript{th} pixel and $\mathrm{d}t_{\ell}\equiv t_{E}$ is the detector's exposure period. Since detectors generate photoelectrons in proportion to the light quanta (\textit{i.e.}, photons) arriving randomly at the $p$\textsuperscript{th} pixel in the $n$\textsuperscript{th} frame, the photoelectron pre-load $C_{n}^{p}$ is Poisson distributed with its rate apportioned by the quantum efficiency $\beta$:
\begin{equation}
    \mathbb{P}\!\left(C_{n}^{p}  \, \middle| \,    \mathit{\mathit{\Phi}}_{n}^{p}) \right. 
= 
    \frac{(\beta\mathit{\Phi}_{n}^{p})^{C_{n}^{p}}}{C_{n}^{p}!}\exp{(-\beta\mathit{\Phi}_{n}^{p}}).
\label{Equation:Poisson(sCMOS)}
\end{equation}
\textbf{Step 2: Analog-to-Digital Conversion}
\par Detectors of sCMOS architecture vary from pixel-to-pixel; such detectors are thusly characterized by pixel-dependent gain $G^{p}$, offset (bias) $O^{p}$, and read-noise variance $(\sigma^{p})^{2}$. For such a detector, photoelectrons $C_{n}^{p}$ are converted into an analog voltage before being digitized to a raw measurement value $w_{n}^{p}$ in ADU\autocite{Huang2013}:
\begin{equation}
    \mathbb{P}\!\left(w_{n}^{p} 
    \, \middle| \,
    C_{n}^{p}
    \right) 
= 
    (2\uppi)^{-\frac{1}{2}}
    \frac{1}{\sigma^{p}}
    \exp\left[ -\frac{\left(w_{n}^{p} - O^{p} - G^{p} C_{n}^{p}\right)^{2}}{2 (\sigma^{p})^{2}} \right].
\label{Equation:MeasurementModel(sCMOS)}
\end{equation}
\subsubsection*{The sCMOS Emission Model}
\par Marginalizing the Poisson distributed photoelectron count $C_{n}^{p}$ out of the Poisson-Gaussian process yields an infinite Gaussian mixture --- the exact likelihood of recording $w_{n}^{p}$ given $\mathit{\Phi}_{n}^{p}$:
\begin{equation}
    \mathbb{P}\!\left(w_{1:N}^{1:P} 
    \, \middle| \,
    \mathit{\mathit{\Phi}}_{1:N}^{1:P}
    \right)
=
    \prod_{n=1}^{N}
    \prod_{p=1}^{P}    
    \sum_{j=0}^{\infty}
      \frac{(\mathit{\mathit{\Phi}}_{n}^{p})^{j}}
                  {j!   \sqrt{2\uppi(\sigma^{p})^{2}}}
      \exp\!\left[
        -\mathit{\mathit{\Phi}}_{n}^{p}
        -\frac{\left(w_{n}^{p}-O^{p}-jG^{p}\right)^{2}}{2(\sigma^{p})^{2}}
      \right].
  \label{Equation:Emission(sCMOS)}
\end{equation}
For detectors of sCMOS architecture, \cref{Equation:Emission(sCMOS)} stands in place of Equation 6 in the main text for any values of $\mathit{\Phi}_{n}^{p}$. With modest illumination (\textit{i.e.}, \(\mathit{\mathit{\Phi}}_{1:N}^{1:P} \ge 5\) photons), the Poisson term in \cref{Equation:Emission(sCMOS)} can be approximated by a Gaussian of variance \(\mathit{\mathit{\Phi}}_{n}^{p}\) without altering the emission model by more than $1\%$\autocite{Mandracchia2020sCMOS}; this approximation yields the closed‑form
\begin{equation}
    \left.
    \mathbb{P}(w_{1:N}^{1:P} 
    \,\middle|\,
    \mathit{\Phi}_{1:N}^{1:P}
    \right)
\approx 
    (2\uppi)^{-\frac{NP}{2}}
    \exp\!\left[-
        \sum_{n=1}^{N}
        \sum_{p=1}^{P}
            \frac{\left(w_{n}^{p} - O^{p} 
            - G^{p} \mathit{\Phi}_{n}^{p} \right)^{2}
        }
        {2\left[ (G^{p})^{2} 
    \mathit{\Phi}_{n}^{p} + (\sigma^{p})^{2} \right]} \right]
    \prod_{n=1}^{N}
    \prod_{p=1}^{P}
    \left\{\left[ (G^{p})^{2} \mathit{\Phi}_{n}^{p} + (\sigma^{p})^{2} \right]\right\}
        ^{-  \frac{1}{2}}
.
\label{Equation:approximateEmission(sCMOS)}
\end{equation}

\section*{Part IV: Supplementary Figures}   
\addcontentsline{toc}{section}{Part IV: Supplementary Figures}
    \subsection*{Motion Models and Anomalousness}
\addcontentsline{toc}{subsection}{Motion Models and Anomalousness}
\begin{figure}[H]
    \centering
    \includegraphics[width=1\linewidth]{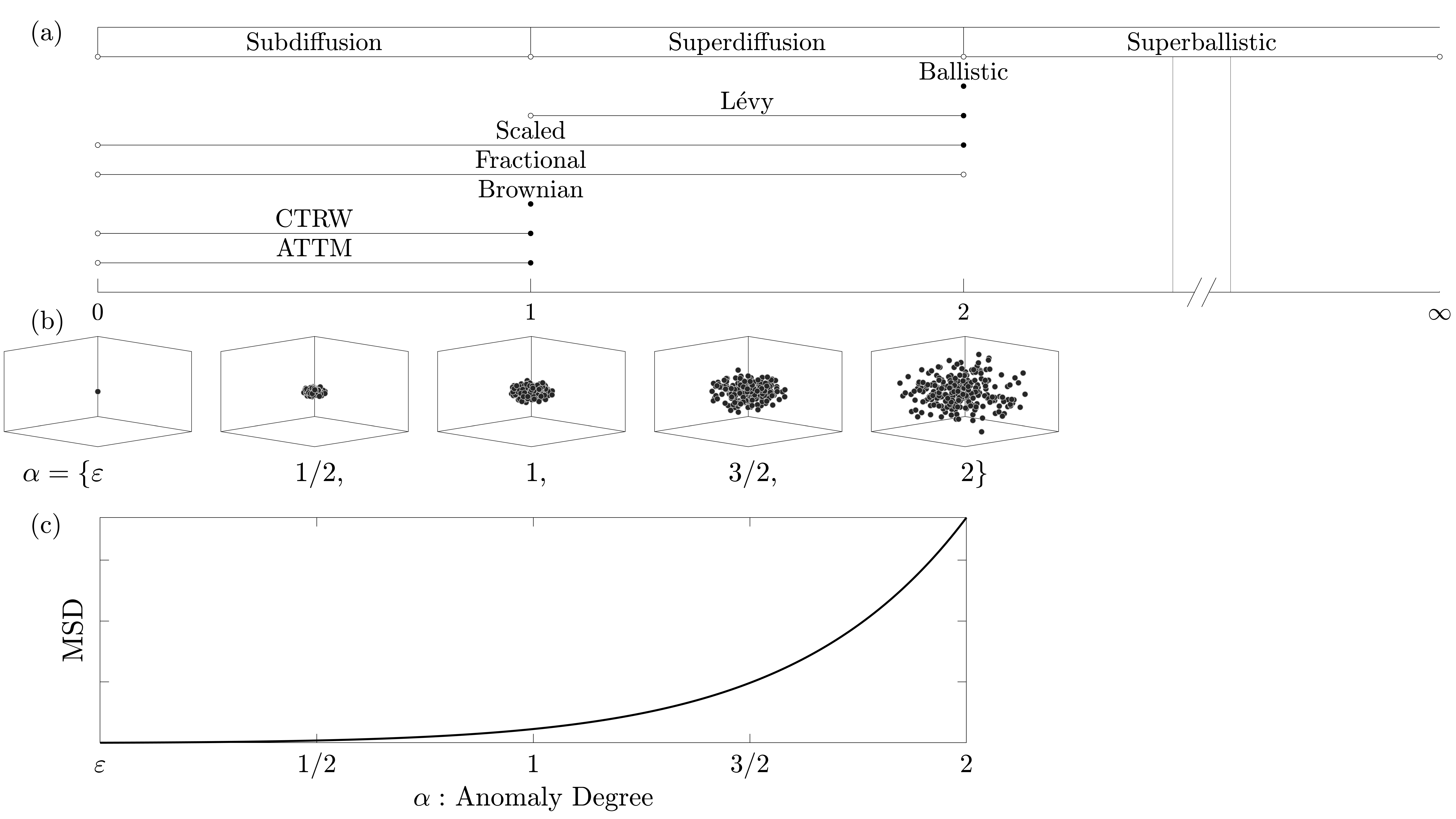}
\caption{Anomalous motion models and the effect of anomalous exponents on scaled Brownian motion (SBM). (a): Here, the anomalous domains and motion models analyzed in this paper are shown within domains of the anomaly degree parameter space. Along the $\alpha$-number line, hollow and solid dots indicate discontinuous and continuous points, respectively. (b): Motional persistence is visualized by the final positions of $250$ diffusing molecules after $10$ steps from the origin, given $\alpha$. Here, $\varepsilon$ denotes a very small number to approximate virtually immobile cases. (c) The ensemble-average mean-square displacement's dependence on the anomaly degree $\alpha$ for the ideal data in panel (b).}
    \label{FigureSI:Panel_Anomaly}
\end{figure}

\par Above, \cref{FigureSI:Panel_Anomaly} shows diffusive regimes of the anomalous exponent $\alpha$. In the subdiffusive (\textit{i.e.}, anti-persistent) domain characterized by stochastic motion in crowded, heterogeneous environments, the annealed transient time motion (ATTM) model~\cite{Manzo2014ATTM} was developed to reproduce trajectories exhibiting patches of localized BM in spatially disordered media, whereas the continuous-time random walk (CTRW)~\cite{Metzler2011CTRW} was designed to simulate diffusion along structural lattices~\cite{Montroll1965} and account for the photoconductivity of amorphous solids~\cite{Scher1975CTRW}.
Non-persistent, pure diffusion --- Brownian motion --- was formulated as a zero-mean Gaussian process with independent, stationary increments~\cite{Wiener1923DifferentialSpace}. In the persistent superdiffusive domain distinguished by quicker, more directed transits, L\'{e}vy distributions~\cite{Levy1937} were used in devising the spatial L\'{e}vy flight~\cite{Mandelbrot1982FractalGeometry, Metzler2007LevyFlights, Metzler2014LevyFlights}, whose divergent $\mathrm{MSD}$ and immediate jumps were remedied by the spatiotemporal L\'{e}vy walk (LW)~\cite{Klafter1982, Klafter1986, Klafter2015LevyWalk}; while the former typifies photon transport through disordered optical media~\cite{Barthelemy2008}, the latter portrays patterns from animal foraging~\cite{Campeau2022} to bacterial chemotaxis~\cite{Huo2021}. Further still, superballistic diffusion is even faster than projectile motion and has been observed in the hydrodynamic flow of electrons through Graphene constrictions. 

\par \cref{FigureSI:Panel_Anomaly} also shows that many anomalous motion models are defined at the Brownian limit ($\alpha=1$), whereat each model is said to converge to $\mathrm{BM}$~\cite{Manzo2021AnDiChallenge}. Included in these models are more generalized Gaussian models encompassing a larger range of the anomalous exponent's parameter space: fractional Brownian motion ($\mathrm{FBM}$) yields subdiffusive, Brownian, and superdiffusive motions~\cite{Mandelbrot1968FBM} best describing stochastic propagation through viscoelastic media~\cite{Metzler2022sBmVERSUSfBm}. Similarly, scaled Brownian motion ($\mathrm{SBM}$) additionally encompasses ballistic diffusion and best represents diffusion coefficients evolving deterministically over time~\cite{Lim2002sBm, Metzler2014sBm} as expected of time-dependent temperature~\cite{Metzler2016UnderdampedSBM} or photobleaching recovery~\cite{Metzler2022sBmVERSUSfBm}. 

\subsection*{Tracking Non-Brownian Motion}
\addcontentsline{toc}{subsection}{Tracking Non-Brownian Motion}
Here, we show accurate tracking of diffusive trajectories across both motion models and the anomalous exponent $\alpha$. Below, $749/750\, (99.9\%)$ of in-frame ground truth positions are circumscribed within our shaded $98\% \, \mathrm{CI}$, missing only a single position for subdiffusive $\mathrm{CTRW}$.
\begin{figure}[H]
    \centering
    \includegraphics[width=1\linewidth]{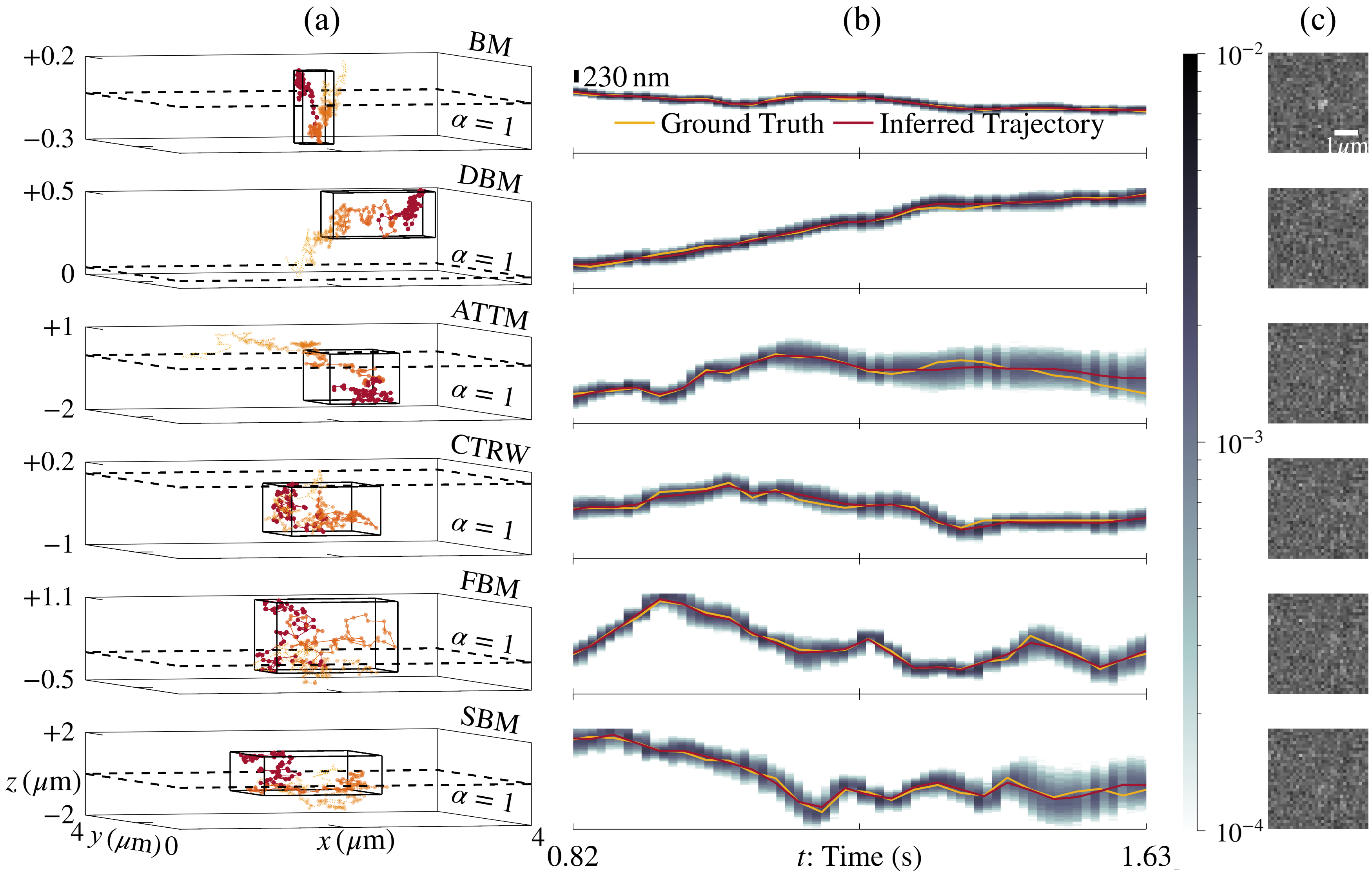} 
\caption
{A likelihood informed by the $\mathrm{BM}$ model accurately tracks particle positions with trajectories generated according to alternate motion models at the Brownian limit. 
(a) Inferred trajectories $\boldsymbol{R}_{1:N}^{1:K}$ are shown for each motion model with opacity, marker size, and color wavelength increasing with time. A dashed line marks the image plane, whereas the solid three-dimensional box encloses the samples shown in the central panel. 
(b) Three-dimensional trajectories along the $\hat{\boldsymbol{x}}$ direction are shown for ground truth (gold) and the mean MCMC sample (red); these trajectories are accompanied by a \qty{230}{\nano\m} scalebar and an associated shading that represents the $98\%$ CI obtained without burn-in over MCMC iterations $i\in[2,6]\cdot10^{3}$. 
(c) The final image $\boldsymbol{w}_{N}^{1:P}$ of each motion model is shown with a \qty{1}{\micro\m} scalebar; these frames have been transposed to align visually with central $x(t)$ plots. The generation of data for this figure is detailed in the \textbf{Forward Models for Data Acquisition} within the \textbf{Methods} section in the main text. A complete list of assigned measurement parameters is provided in \textbf{Table 3} in the main text.}
\label{FigureSI:Results_Brownian}
\end{figure}
\begin{figure}[H]
    \centering
    \includegraphics[width=1\linewidth]{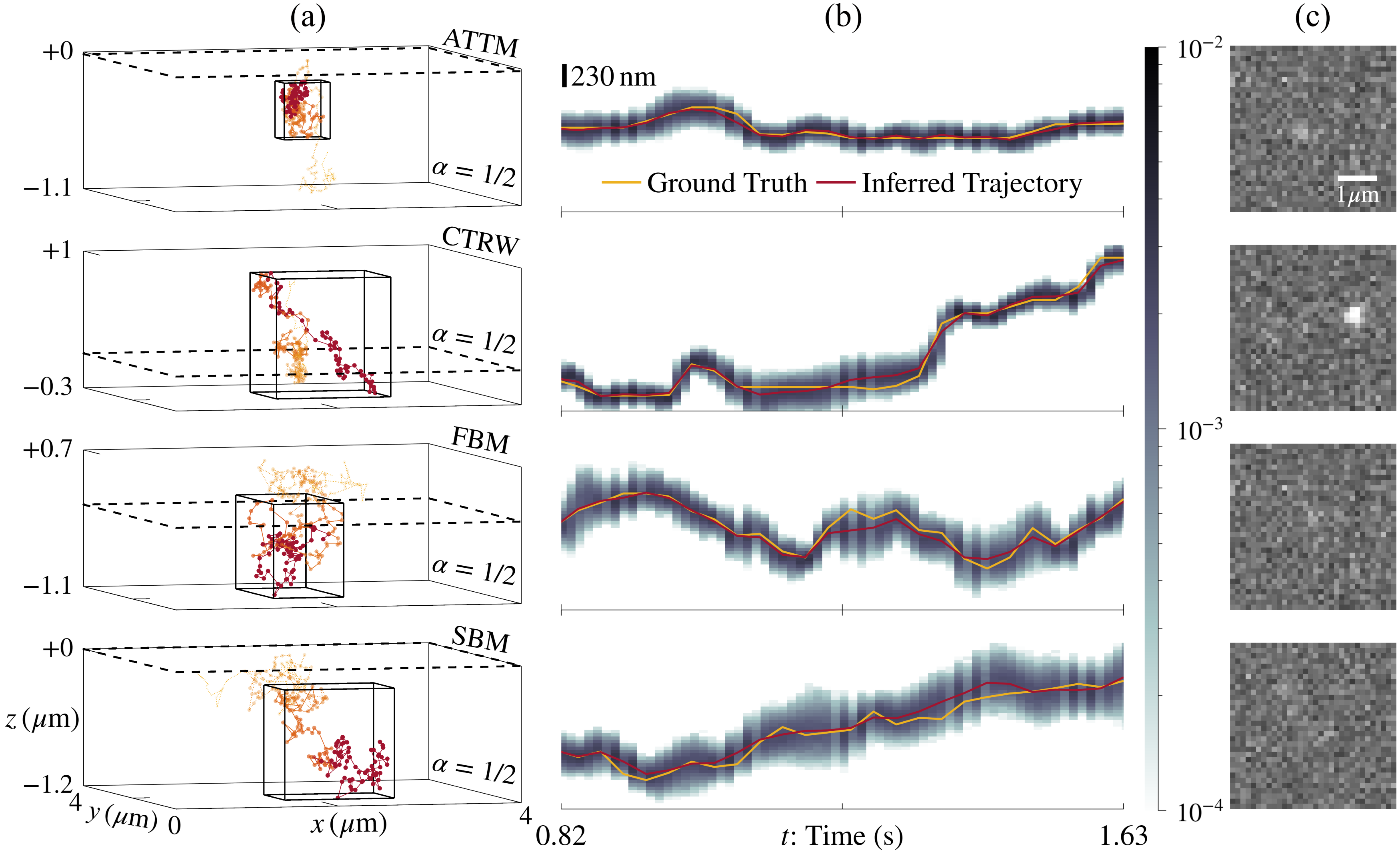} 
\caption
{A likelihood informed by the $\mathrm{BM}$ model accurately tracks particle positions with trajectories generated according to subdiffusive motion models. The figure's layout is identical to \cref{FigureSI:Results_Brownian}. A complete list of assigned measurement parameters is provided in \textbf{Table 3} in the main text.} 
\label{FigureSI:Results_Sub}
\end{figure}
\begin{figure}[H]
    \centering
    \includegraphics[width=1\linewidth]{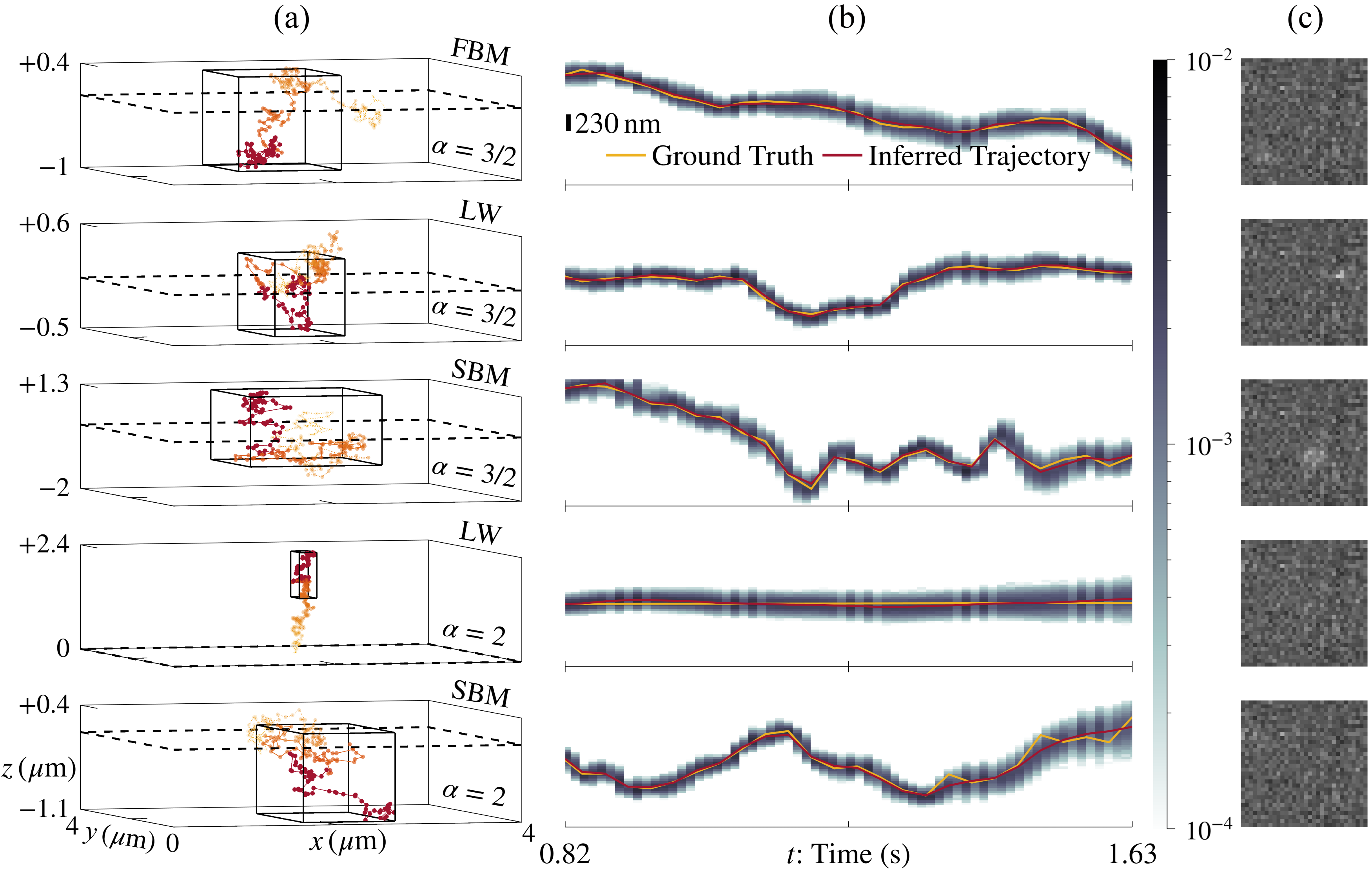} 
\caption
{A likelihood informed by the $\mathrm{BM}$ model accurately tracks particle positions with trajectories generated according to superdiffusive and ballistic motion models. The figure's layout is identical to \cref{FigureSI:Results_Brownian}. A complete list of assigned measurement parameters is provided in \textbf{Table 3} in the main text.} 
\label{FigureSI:Results_Persistent}
\end{figure}

\section*{Part V: Supplementary Tables}
\addcontentsline{toc}{section}{Part V: Supplementary Tables}
\begin{table}[H]
    \centering
    \begin{tabular}{|c|c|c|} \hline 
         Method (or Team) Name&  Language& Mode of Failure\\ \hline\hline
         Teams \{A\autocite{AnomalousUnicorns1992, AnomalousUnicorns2022}, C:G\autocite{Gratin, DeepSPT2016DeepResidualLearning, DeepSPT2016ScalableBoosting, TeamF, RANDI}, J:K\autocite{TeamJ2018Networks, TeamJ2019DeepLearning, TeamK}, N:O\autocite{TeamN2018, TeamJ2019DeepLearning, TeamN2019, TeamO2019, TeamO2020Impact, TeamO2020}\}&  Python& Library Deprecations\\ \hline 
         \textit{AnDi-ELM}\autocite{ELM}&  MATLAB& Instructionless\\ \hline 
         \textit{NOBIAS}\autocite{NoBIAS}&  MATLAB& Defunct Example\\ \hline\hline
         Teams \{G\autocite{WadNET}, H\autocite{TeamH} \}&  &   Only Infers $\alpha$\\ \hline 
 Teams \{A\autocite{AnomalousUnicorns1992, AnomalousUnicorns2022}, D\autocite{DeepSPT2016DeepResidualLearning, DeepSPT2016ScalableBoosting}, H\autocite{TeamH}, I, K\autocite{TeamK} \}& &1-Dimensional\\\hline\hline
 \textit{Gratin}\autocite{Gratin}, \textit{AnomDiffDB}\autocite{TeamJ2018Networks, TeamJ2019DeepLearning}& &2-Dimensional\\\hline
    \end{tabular}
\caption{The majority of methods for decoding anomalous diffusion are either currently unusable or only useful for analyzing one- or two-dimensional trajectories.}
\label{Table(Fails)}
\end{table}

\printbibliography[heading=bibintoc,title={References (Supplementary)}]
\end{refsection}

\end{document}